\documentclass[1p, 10pt]{elsarticle}
\usepackage[normalem]{ulem}

\usepackage{graphicx}

\usepackage[hyphens]{url}
\usepackage[hidelinks]{hyperref}
\hypersetup{breaklinks=true}
\usepackage{geometry}

\usepackage{amssymb}
\usepackage{wasysym}
\usepackage{lscape}
\usepackage{xcolor}
\usepackage{enumitem}
\usepackage{textcomp}
\usepackage{float}

\usepackage{array}
\usepackage{afterpage}

\usepackage{xcolor}
\usepackage{textpos}
\usepackage{multirow}
\usepackage{tablefootnote}
\usepackage{longtable}
\usepackage{pdflscape}

\usepackage[yyyymmdd,hhmmss]{datetime}

\newcounter{magicrownumbers}
\newcommand\rownumber{\stepcounter{magicrownumbers}\arabic{magicrownumbers}}

\usepackage[font=small]{caption} 

\begin{document}
\begin{frontmatter}

\title{Understanding Cyber Threats Against the Universities, Colleges, and Schools}

\author{Harjinder Singh Lallie*}
\author{Andrew Thompson \textdaggerdbl}
\author{Elzbieta Titis*}\ead{elizabeth.titis@warwick.ac.uk}
\author{Paul Stephens\textdagger}
\address{*Cyber Security Centre, WMG, University of Warwick, Coventry, CV4 7AL, \textdaggerdbl Kagool, 5th Floor, One Friargate, Station Square, Coventry, CV1 2GN, \textdagger Canterbury Christ Church University,  North Holmes Road, Canterbury, Kent, CT1 1QU}

\begin{abstract} 
Universities hold and process a vast amount of valuable user and research data. This makes them a prime target for cyber criminals. Additionally, universities and other educational settings, such as schools and college IT systems, have become a prime target for some of their own students -- often motivated by an opportunity to cause damage to networks and websites, and/or improve their grades.

This paper provides a focused assessment of the current cyber security threat to universities, colleges, and schools (`the education sector') worldwide, providing chronological sequencing of attacks and highlighting the insider threat posed by students. Fifty-eight attacks were identified, with ransomware being the most common type of external attack, and hacking motivated by personal gain showing as the most common form of internal attack. Students, who have become a significant internal threat by either aiding or carrying out attacks are not a homogeneous group, as students may be motivated by different factors, therefore calling for targeted responses. Furthermore, the education sector is increasingly reliant on third party IT service providers meaning attacks on third parties can impact the university and its users. There is very little research analysing this problem, even less research analysing the problem in the context of schools. Hence this paper provides one of the first known assessment of the cyber attacks against the education sector, focusing on insider threat posed by students and offering recommendations for mitigating wider cyber threats.

\end{abstract}

\begin{keyword}
cybercrime \sep UK universities \sep cyber attack timeline \sep insider threat \sep human factors \sep cyber security awareness

\end{keyword}

\end{frontmatter}

\section{Introduction}\label{Introduction} Universities hold critical research data as well as information on more than 3 million staff and students and millions of alumni \citep{hesa2021staff, hesa2021student}. University networks are extraordinarily complex with tens-of-thousands of users logging on daily using a variety of personal devices, hosting a range of cyber security protection. In addition, users log on from a wide range of locations, and have a wide range of motivations for accessing the data that they do. University websites are designed with an inherent level of openness revealing, amongst other data, course tutor, support staff, and research staff details. This is a large and diverse surface area potentially encompassing a huge range of vulnerabilities. 

This technical landscape creates easier opportunities for perpetrators to launch cyber attacks, particularly targetting people in jobs that have access to strategic or confidential data \citep{lee2020security}. Universities have become popular cyber crime targets for various adversaries such as state-sponsored actors, criminals, and insiders \citep{bongiovanni2019least, Singar2020role}. At the time of writing, spear-phishing and social engineering remain the main external attack vectors against universities, with ransomware increasingly becoming the greatest single cause of disruption \cite{NCSC}. The types of data targeted include emails, bulk personal information on staff and students, technical resources, and sensitive research and intellectual property \cite{NCSC}. While the effects of cyber crime are perhaps most disruptive for universities, state-sponsored espionage is likely the cause of greater long-term damage, including damage to the value of research and the UK’s knowledge advantage \cite{NCSC}. 

Other than a report by JISC \citep{JISC2022}, there is a lacuna of research surrounding cyber attacks on, and the cyber security posture of, educational establishments such as universities, colleges, and schools. The present study draws on publicly available data to outline the range of prevalent attacks against the education sector.

In the UK, `college’ usually refers to further education (FE) attended by students at the age of 16, after school, and before university. Students from the age of 16 can either choose a state sixth form college or a college of further education as an alternative to private education, both types preparing them for entry to an university. This is in contrast to the US terminology in which `college’ is often used generally to refer to university, or specifically undergraduate institutions in universities, such as Yale College or Harvard College. We define `school' according to the Education Act 1997 as an educational institution which is outside the FE sector and the HE sector, and is an institution for providing (a) primary education, (b) secondary education, or (c) both primary and secondary education. In this paper, we refer jointly to schools and colleges as `schools' to distinguish them from universities. 

Our key contribution includes a novel timeline of key cyber attacks, providing chronological sequencing of attacks to serve as a platform to align with current literature, as well as offering the foundation for others to build on. We also compile a list of attacks carried by students against their home schools and universities to help frame the discussion of insider threat.

The paper is organised as follows: relevant cyber attack and cyber crime literature is reviewed in Section \ref{Background} (including discussion of human factors in academic settings, most common threats faced by universities worldwide, and university cyber defence weaknesses), followed by a discussion of our methodology for creating a cyber attack timeline in Section \ref{Method}. We then discuss results in Section \ref{Results} followed by presenting a broader reflection on the attacks in Section \ref{Discussion}, the latter including discussion of insider threat and human factors moderators, impact and mitigation, and the potential limitations of the work before concluding and proposing future work in Section \ref{Conclusions}.

\section{Background}\label{Background} This section analyses published research into cyber-attacks on universities. We were not able to identify any meaningful research into cyber security, and specifically cyber attacks, against schools and/or colleges. This outlines a clear research gap which the present contribution aims to address.

Cyber attacks against universities have been increasing in recent years \cite{csbs, teiss2020warwick, bbc2020blackbaud}, with the average cost of addressing a cyber attack amounting to £620,000 in 2021 \cite{cost}. The most significant rise in cyber attacks has come in the form of ransomware attacks \cite{ncsc2020septalert}. JISC provide example costs associated with recovery from cyber attacks. A 2019 university phishing attack resulted in a cost of £65,000 related to staff response time and ``significant legal costs''; a 2020 phishing campaign against a UK college led to recovery costs of c£30,000; a 2020 password compromise led to a response cost of £40,000; another 2020 password compromise led to a compromise of users in other universities leaving around 1,000 accounts compromised and around 80 days of response effort. These figures do not include the opportunity cost created by the redeployment of staff and resources which might not otherwise have expected to have been applied, nor the time lost for end users \citep{JISC2020}.

University infrastructures are growing, student numbers are increasing, and investment in buildings and resources is rising across the world \cite{loeppky}. Recently, as COVID-19 brought greater reliance on digital systems and networks \cite{lallie2021cyber}, the pandemic also expanded the attack surface with most businesses and institutions moving services online \cite{NCSC2021}.

The impact of a cyber attack extends across multiple dimensions and beyond the confines of the university \cite{jones}. For example, were attackers to steal critical research through a cyber attack, the researchers would lose the opportunity to publish the research, the university could become less valuable to investors, and funding may be impacted \cite{jones}. Such an attack could impact other sectors too. So if the research was of a particularly sensitive nature to bodies such as the military or national defence, this could pose a grave threat to the whole country \cite{jones}. The extended impact is exemplified by a May 2020 cyber attack where a third-party service experienced a ransomware attack, which impacted several universities both in terms of data availability and confidentiality (exfiltration) \citep{JISC2020}. Awareness, training, and certification are three of five solutions posited in the first version of the JISC report as mechanisms to reduce the likelihood of a cyber attack against a university, the other two being strong leadership and adherence to technical basics \citep{JISC2020}. Version 2 of the JISC report reflects more recent incidents and associated guidance, and stresses the importance of cyber insurance \cite{JISC2022}.

Although certification schemes such as Cyber Essentials exist to enable organisations to improve their cyber security posture, many UK businesses and charities are unaware of these schemes \cite{csbs}, resulting in poor cyber security practices. In 2017, the average self-rating of a university's cyber security posture was 5.8 out of 10, rising to 6.5 for those that had obtained the Cyber Essentials certification \citep{chapman2018severity}. More recently, the JISC Posture Survey \citep{JISC2020} revealed a perception of low cyber security posture amongst HEI and UK college stakeholders with a mean score of 6.4/10 and 7/10 respectively and just 10\% and 36\% scoring their institution more than 8. In the following sections we discuss the value of training and reporting to ongoing efforts in the fight against cyber crime, we also present common attacks faced by universities globally, with special considerations for internal threats posed by students.

\subsection{Training and reporting}

Communication is important in achieving an acceptable cyber security culture, which could prepare universities to battle ongoing attacks better than big funding or sophisticated technical systems \citep{pavlova2020enhancing}. However, a survey of UK universities revealed that: a) only 54\% of UK university staff had received cyber security training, universities were only spending an average of £7,529 on cyber security training for staff in 2019; b) 51\% of universities proactively issued cyber security training and information to students; and c) 37\% only offered support to students that requested it whereas 12\% did not offer any kind of cyber security guidance or training to their students \citep{Redscan}. Table \ref{Table-PreviousStudies} shows a sample of studies that have examined cyber security awareness and its relationship with various human factors involving students. 

Overall, students had a good cyber security intuitive judgment, but their overall competence differed in regards to various aspects of cyber security \cite{yan2018finding}, especially in relation to distinguishing between genuine and spear-phishing emails \cite{butavicius2016breaching}. There is indeed a need and a space for improvements regarding cyber awareness \cite{zwilling2019deal}, but student engagement with such initiatives is limited \cite{potgieter2019awareness}. \cite{zwilling2019deal} developed a practical cyber awareness education framework (CAEF) to assist in building the tailored syllabus, showing a positive association between cyber security awareness and cyber security behaviour \citep{zwilling2020cyber}. The large number and wide variety of users at universities may pose challenges in terms of tracking who has completed training and developing a training regime that adequately covers appropriate assets for individuals \cite{lane2007information}; moreover, these challenges may be further exacerbated by high staff turnover \cite{bongiovanni2019least}.

\begin{table*}[!hbp] 
\scriptsize
\begin{tabular}{|p{0.5cm}|p{2cm}|p{1cm}|p{2.2cm}|p{5.9cm}|} 
\hline
\textbf{Ref} & \textbf{Study Population} & \textbf{Sample*} & \textbf{Human Factor Moderators} & \textbf{Results} \\ 
\hline

\cite{butavicius2016breaching} & Unnamed large South Australian university & 121 & Cyber security behaviour, knowledge of cyber security, social engineering strategy, decision making style & Users struggled to distinguish between genuine and spear-phishing emails. When detecting phishing and spear-phishing emails, users performed the worst when the emails used the authority principle and performed best when social proof was present. Users who were less impulsive in making decisions generally were less likely to judge a link as safe in the fraudulent emails. \\ 
\hline

\cite{yan2018finding} & Unnamed northeastern United States of America public university & 462 & Cyber security behaviour, knowledge of cyber security, intuitive or rational judgement & The average percentage correct of cyber security judgment was about 65\%. Three groups of ordinary users are low cyber security judgement, moderate cyber security judgement, and high cyber security judgement. Students also differed in their judgment competence on different aspects of cyber security and their rational judgment underperformed their intuitive judgment. \\ 
\hline

\cite{broadhurst2019phishing} & Australian National University, Australia & 138  & Cyber security behaviour, knowledge of cyber security, social engineering strategy, gender, international or domestic, primed to cyber attacks & Although tailored and individually crafted email scams were more likely to induce engagement than generic scams, differences were not significant. International students and first year students were deceived by significantly more scams than domestic students and later year students. \\ 
\hline

\cite{potgieter2019awareness} & Central University Technology, South Africa & 43 & Level of social media use, cyber security training history & There apears to be a reluctance amongst students to engage with cyber security awareness (CSA) initiatives that are available. Students engaged with social media platforms at least once a week with Facebook and YouTube the most popular. They also used communication media like websites to pursue material about CSA. \\ 
\hline

\cite{zwilling2019deal} & International School for Business and Social Studies, Slovenia & 35  & Cyber security behaviour, knowledge of cyber security, computer skill, willingness to attend training & There is a need and a space for improvements regarding cyber awareness. Detailed recommendations and cyber awareness education framework (CAEF) proposed for curricula changes at each level of the evaluated education systems and practice. \\ 
\hline

\cite{aljohani2020measuring} & Fahad Bin Sultan University, Saudi Arabia & 189  & Cyber security behaviour, knowledge of cyber security & Students’ awareness was average with no difference in awareness levels between male and female students. \\ 
\hline

\cite{redman2020improving} & University of New South Wales, Australia & 158 & Cyber security behaviour, knowledge of cyber security, training style preference & Students struggled to remain engaged and found the labs were pitched at a high-level. Applying proven teaching methods, following internationally recognised frameworks and recommendations, improved the level of engagement. \\ 
\hline

\cite{zwilling2020cyber} & Ariel University, Israel; Lublin University, Poland; International School for Business and Social Studies, Slovenia; unnamed Turkish universities & 459** & Cyber security behaviour, knowledge of cyber security, computer skill, willingness to attend training, cyber security training history, county of education, gender & Users possess adequate cyber threat awareness but apply only minimal protective measures usually relatively common and simple ones. Higher cyber knowledge is connected to the level of cyber awareness, beyond the differences in respondent country or gender. Awareness is also connected to protection tools, but not to information users were willing to disclose. There are also differences between the countries that affect the interaction between awareness, knowledge, and behaviors. \\
\hline

\multicolumn{5}{l}{* All the studies used convenience sampling except \cite{aljohani2020measuring} who used random sampling.}\\
\multicolumn{5}{l}{**89 Israeli students; 182 Polish students; 35 Slovenian students; 153 Turkish students.} \\
\hline
\end{tabular}
\caption{Summary of previous studies that looked at cyber security awareness at universities, including results and significant predictors examined.}
\label{Table-PreviousStudies}
\end{table*}

In addition to training, reporting may be of particular value to ongoing efforts in the fight against cyber crime. However, public disclosure of sustained attacks does not appear common and there may be significant levels of under-reporting, with only three in ten incidents being reported in 2020 \citep{nfib2021cybercrime}. Full disclosures, such as that by the University of Sunderland, which sustained a cyber attack in 2021 and chose to share this information publicly soon after, are quite rare \cite{loeppky}. 

A wide range of reasons may explain low reporting by victims, such as not knowing of being defrauded, feeling partly responsible, or embarrassed, the attitude of statutory bodies, and overall confusion. Since enhanced cyber security leads to higher identification of attacks, less cyber mature organisations may be more likely to under report \cite{csbs}. Moreover, because data breaches dent university finances and reputations, it is also possible that universities choose not to report due to balancing the perceived costs of reporting against the expected benefits. Indeed, financial and reputational concerns often prevent financial businesses to report cyber crime from the fear of losing trust in the company and its products \cite{Lagazio}. The type of cyber crime and its impact may be relevant to the reporting decision for businesses, whereas having cyber security insurance may not \cite{Kemp}; the relative importance of these factors to reporting cyber crime for HEIs is currently unknown.

\subsection{Common attacks faced by universities}

A variety of threats targeting both students and staff have been recognised in the literature, including phishing, DDoS attacks, ransomware, and terrorist and internal threats.

\paragraph{Phishing} Phishing remains one of the most popular attack techniques, with some universities receiving millions of phishing emails each year \citep{redscan2020analysis}. Spear-phishing may lead to a speedy compromise of university networks, providing access to valuable data including personal data, finance systems, and research networks \citep{chapman2019safe, torjesen2019universities}. 

Cyber threat actors often leverage societal events to take advantage of human factors. A good example of this was the plethora of cyber crime related reports which were linked to the COVID-19 pandemic \citep{lallie2021cyber}. Similarly, the start of an academic year increases the prevalence of phishing emails that offer free grants or requests for bank details to be updated so that loans can be paid \citep{chapman2019safe}. One example includes the phishing attack carried against the University of New South Wales, in which the perpetrator compromised the Facebook account of the university, encouraging followers to consider alternative top universities around the time of an open day \citep{bagshaw2015unsw}.

Sophisticated campaigns, such as \textit{Silent Librarian}, have also compromised the credentials of academic staff and students through highly-targeted socially engineered spear-phishing emails \citep{chapman2019safe, malwarebytes2020silent}. These credentials have been used to access important research or proprietary academic information \citep{tabansky2018iran, seals2020silent}. Other campaigns, such as \textit{Stolen Pencil} involved spear-phishing attacks distributing  malicious Trojans \citep{zurkus2018stolen,mitre2020stolen}, aimed specifically at academics in certain disciplines such as biomedical engineering \citep{netscout2020stolen}.

\paragraph{DDoS attacks} In 2016, 63 UK universities had suffered from DDoS attacks \citep{chapman2018severity}, whilst in 2019 DDoS attacks continued to be on the rise, with a successful attack on the University of Edinburgh making headline news \citep{chapman2019safe}. In 2021, there was a 102\% increase in such attacks targeting universities, colleges, and schools with a DDoS attack occurring every three seconds \citep{netscout2022}.

\paragraph{Ransomware} Ransomware attacks against schools,
colleges and universities in the UK have been on the rise since August/September 2020, with significant spikes in February 2021 and again in May/June 2021 \citep{NCSCalert}. Prior to that it was estimated that a third of UK universities were subjected to ransomware attacks between 2010 and 2020 \citep{coker2020ransomware}. As of 2021, ransomware has become the most significant cyber threat for businesses and small and medium enterprises (SMEs) in the UK \cite{NCSC2021}, having an impact comparable to state-sponsored espionage \citep{NCS2022}. Globally, it has been estimated that in 2022, 5\% of all ransomware attacks were directed at education sectors, with the average data breach cost in education at 3.86 million dollars \citep{ibm2022}.

Although law enforcement do not encourage, endorse, nor condone the payment of ransom demands, several universities have reported paying cyber criminals to unlock their systems, such as Maastricht University that paid nearly €200,000 of bitcoin to regain access to research and recover its commercial operations \citep{loohuis2020maastricht}, and the University of California, San Francisco that paid attackers \$1.14 million to recover hacked data from its School of Medicine \cite{gt}.

However, payment of the ransom does not guarantee a return to full system restoration, as in the case of Regis University, which had day-to-day operational disruption for months after paying a ransom \citep{hernandez2020regis}. Ransomware may also present an existential threat to colleges and smaller universities that may not be able to withstand the financial impact of paying a high ransom. It is therefore essential for universities to enact a range of measures to help recover from ransomware attacks such as keeping offline backups of their most important files and data, in addition to utilising a defence-in-depth strategy for mitigating malware and ransomware attacks \cite{NCSC2}. 

Universities are prone to cyber attacks from a wide variety of actors which include state, terrorist, hacktivist, and even their own students. 

\paragraph{State, terrorist, and hacktivist actors} Terrorist organisations can also compromise universities' websites to display messages promoting their causes, as in the case of Tsinghua University in China having its website compromised by putting a photograph and audio in support of holy war \citep{chen2016islamic}, and University of Botswana having its logo replaced by an image of the hacktivist group Anonymous \citep{sunday2017it}. Cyber criminals have also targeted Australian and American university websites, in some cases inserting hyperlinks and malicious code directing students to ``essay mill'' sites to seemingly legitimise professional contract cheating services \citep{qaa2021emerging}, or to fake essay contests that harvest the work of students for resale \citep{mckie2021australia}. 

State-sponsored cyber warfare instances include the Australia National University allegedly attacked by the Chinese state \citep{mcgowan2019china}, Korea University allegedly attacked by the North Korean state \citep{korea2012nk}, and Bar Ilan University allegedly attacked by Iran using a \textit{wiper} attack \citep{ziv2021cyber, ehrlich2021new}. In 2022, the education sector were the target of nation-state actors 14\% of the time \citep{microsoft}, with North Korea being allegedly responsible for 23\% of all the attacks \citep{microsoft}.

Other malware attacks have led to disruption to universities, and institutions having to implement measures to protect their students and staff. For example, the University of Giessen had to resort to physically handing out new email passwords to its 38,000 students and staff after their network was inaccessible for several weeks, a process that took a week \citep{holmes2019giessen}. Similarly, after a malware attack on a software application used for financial management at the University of California, Berkeley, the university provided its impacted students and staff with a year of free credit monitoring and identity theft insurance, along with resources to assist them in monitoring their accounts for any suspicious activity \citep{gilmore2016campus}.

\paragraph{Internal threat} Having a unique level of institutional access, students may present a particular type of threat either as targets of phishing attacks, or for deliberate malicious reasons such as viewing and altering grades, playing pranks, testing their own hacking abilities, or carrying revenge. The literature highlights cases of student grade manipulation, including attacks at Austin's Business School \citep{Singar2020role} and instances within Kenyan universities where the outstanding fee balances for some students had been altered \citep{maranga2019emerging}. Moreover, at the University of Alberta, a student compromised over 3,300 accounts \citep{fitzpatrick2017student}, whereas an insider at Rutgers University took down the institution's central authentication server that maintained the gateway portal to deliver assignments and assessments \citep{bisson2018rutgers}.

In this paper we examine cyber crime that has taken place within the education sector. We investigate external and also internal student generated cyber attacks. There is a distinct lack of academic literature in this domain, as many studies that have researched cyber crime at universities are from an international context, usually from a specific country without international comparisons. For this reason, our research required greater reliance on data and information provided by cyber security companies, news sources, and government organisations than traditional academic outlets. In the follow up section, we describe our data and methodology in more detail.

\section{Timeline creation methodology}\label{Method} There is very little academic data available on the problem of internal and external cyber threats against the university, college, and schools sector, therefore our literature search relied largely on new sources. Our data gathering method was similar to that used in \cite{lallie2021cyber}. For each cyber threat, we identified multiple sources. A mixture of traditional repositories (e.g., Web of Science, Scorpus) and other reputable sources were considered, the latter including news outlets (Reuters, BBC), blog articles, security company reports, social media posts, and search engines (Google, Bing). Occasionally, some elements of the intelligence contradicted each other. We had to reason with this and only use reliable data following a typical cyber-intelligence operation. Only articles in English were included in the analysis; we ignored cases where we were not able to attribute the attack. 

Cyber security news sites are a useful source of information on cyber attacks. In our opinion, some of the reputable sources of information on cyber attacks include: computerweekly.com, cybersecurity-insiders.com, cyware.com, infosecurity-magazine.com itgovernance.co.uk, itpro.co.uk, recordedfuture.com, securityweek.com, thehackernews.com, theregister.com, threatpost.com, and zdnet.com. The Google ``site:'' function was used to search these websites. As these sites focus on cyber security, the search terms ``university'', ``universities'', and ``Higher Education'' could be used to find any news items that may mention a cyber attack on a university. 

A variety of keywords were used when collating reports of cyber attacks, including combination of keywords, e.g., ``university'' and ``cyber attack''. One of the challenges has been that of the use of inconsistent terminology to refer to important concepts within the domain (e.g., perpetrator, hacker, attacker to mean the same thing), as previously highlighted in \cite{lallie2021cyber}. The same problem exists in terms of categorising motivation of perpetrators, as many different categories have been proposed \cite{cheng2020individual, Rogers}. The review of reports was performed up to 2022, providing a basis to be built upon in future studies, both for future attacks and study gaps.

Twitter was used to locate any news on university cyber attacks that had received less publicity. This was done using the advanced search term ``university cyber attack until:2022-12-31 since:2015-01-01 -filter:replies''.

Despite our best efforts to discover what we could about each cyber-attack, some attacks remained shrouded in silence. E.g. the cyber attack on University of Hertfordshire (number 51 in Table \ref{Table-CombinedCyberAttacks}) was believed to be ransomware, but no information was released to confirm this \citep{coker2020ransomware, speed2021hertford, scroxton2021hertfordshire}. Similarly the attacks on Glasgow Caledonian University (row 52) and University of Sunderland (row 55) did not reveal full details.

Although the names of perpetrators in the case studies we refer are in the public domain, other than in a few famous cases, we have anonymised names within this report.

\afterpage{%
\newgeometry{left=6mm,right=6mm}
    \begin{landscape}
    \begin{figure*} [!b]
        \includegraphics[width=1\linewidth]{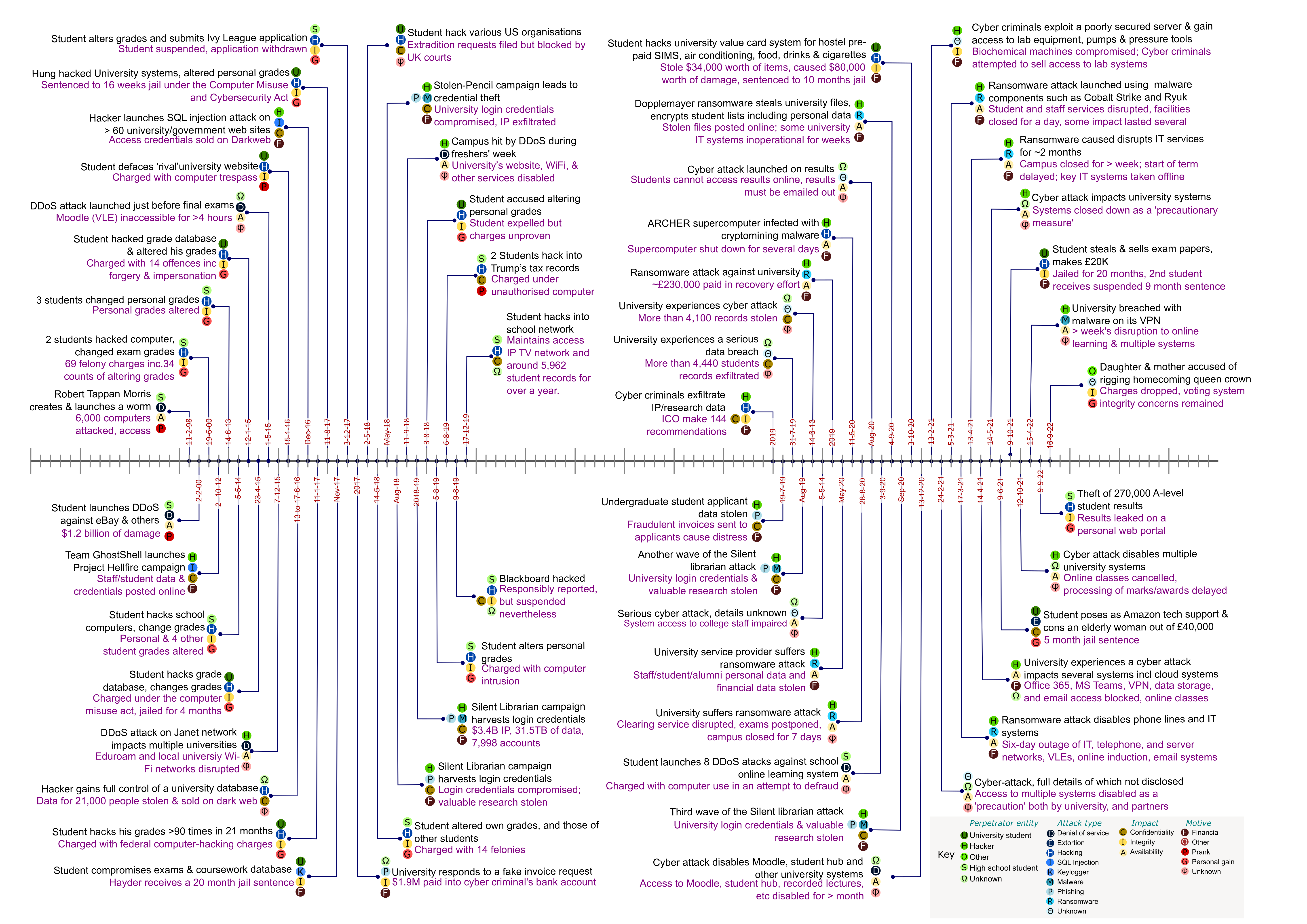}
        \caption{Timeline of Key Cyber-Attacks on UK Universities}
        \label{Figure-UK-Atttack-Timeline}
    \end{figure*}
    \end{landscape}
\clearpage
\restoregeometry

}

\section{Results and analysis of cyber-attacks and associated risks}\label{Results}

In this section, we examine the cyber attacks on universities, colleges, and schools in further detail. The data provided herein outlines the attack date, type (e.g., physical attack or DDoS), victim, description and impact. The discussion is based on a chronological timeline provided in Table \ref{Table-CombinedCyberAttacks} and supported by Figure \ref{Figure-UK-Atttack-Timeline}.

Figure \ref{Figure-UK-Atttack-Timeline} provides a detailed temporal representation of the chain of key cyber attacks, aiding an understanding of how universities were exploited and impacted by cyber crime. The timeline is presented chronologically. Each cyber attack is coupled with four icons which represent, from top to bottom, the type of perpetrator, the type of attack, the impact of the attack, and the motive. Each of these is then further subdivided as follows:

\begin{itemize}[noitemsep]
    \item \textbf{Perpetrator entity:} hacker, other, high school student, unknown
    \item \textbf{Attack type:} denial of service, extortion, hacking, SQL injection, keylogger, malware, phishing, ransomware, unknown
    \item \textbf{Impact:} confidentiality, integrity, availability
    \item \textbf{Motive:} financial, other prank, personal gain, unknown
\end{itemize}

These are not conclusive sub-categories. For example, there are many other attack types, however, we have only included categories revealed in the research. 

A key at the bottom right of Figure \ref{Figure-UK-Atttack-Timeline} explains the colour coding and the categories.

The data presented in Figure \ref{Figure-UK-Atttack-Timeline} is derived from Table \ref{Table-CombinedCyberAttacks} which summarises cyber attacks against education institutions. The data presented includes attack type, date, and additional description of attack, attribution (name of offender), and impact (e.g., altered grades or prank).

Fifty-eight cyber attacks were identified, with the oldest and the most recent ones taking place in February 1988 and September 2022, respectively. Universities from different groups have been compromised, with highest numbers of reported attacks showing in 2019 (thirteen attacks), 2020 (nine attacks), and 2021 (ten attacks).

Each cyber attack was categorised and the impact was also outlined. So, `D.A' means a denial of service attack which resulted in loss of availability. Where there is data missing in the attack type/impact column, this is because the data was not available to the authors. A key at the bottom of the table explains this categorisation system. We can presume that all publicly known attacks resulted in a loss of reputation and financial loss. Although we have categorised the cyber attacks into a single overarching category, in reality, the cyber attacks may have involved multiple attack elements. So, a ransomware attack may have involved phishing and a DoS attack before the ransomware was activated. Our data does not gather this level of granular detail. Although in some cases the impact of an attack may have both exfiltration and loss of availability, this was not specified in any of the attacks that we analysed. 

The data in Table \ref{Table-CombinedCyberAttacks} reveals 2 injection attacks, 4 denial of service attacks, 6 phishing attacks and 7 ransomware attacks. Aside from reputational and financial impact, loss of availability was the more dominant impact with 21 of the cases experiencing this, and 12 institutions admitting to the exfiltration of data.

Prior to 2015, we found only single instances of reported attacks each year, however, good quality reporting on cyber attacks before 2012 was hard to uncover due to poorer legislation. As legislation has tightened, and organisations are under more stringent pressure to report cyber security breaches, the quality of data became better.

\subsection{Analysing the Insider Threat}

Our analysis depicted 28 cases of insider attacks by both university (50\%, n=14) and school students (50\%, n=14), majority of which resulted from hacking (61\%, n=17) and where the intention was to alter marks (28\%, n=8) or playing pranks (21\%, n=6). We also noted some differences between the two groups of students in terms of motivation and types of attack carried; for example, school students carried more DDoS attacks and were more likely to play pranks, whereas university students were more financially motivated.

Table \ref{Table-CombinedCyberAttacks} shows that the majority of attacks by university students was for the purpose of altering grades (35\%, n=5), followed by financial gain (21\%, n=3) and playing prank (14\%, n=2), however, students were motivated by various illegal gains, including hacking, legitimate challenges set by tutors, copyright theft, and personal gain. When comparing results with school students, the motivation to carry cyber attack was less varied and mostly perpetrated for the purpose of altering grades (28\%, n=4), playing pranks (28\%, n=4), and receiving personal benefit (n=3, 21\%). 9 cases of grade manipulation among both group of students indicate that improving grades is important motivational driver for cyber attacks against educational institutions \cite{DoF2022}.

The majority of attacks carried by university students resulted from hacking (71\%, n=10), including 1 harmless hacking, and 4 other attacks were enabled by either virus, malware, copyright theft, web site defacement, and phone fraud. Similarly attacks by school students were mostly the result of hacking (57\%, n=8), however, there were more DDoS attacks in the mix (21\%, n=3), and other unique types of attacks included single instances of deception, bug hunting, and cyber attack.

\section{Discussion}\label{Discussion} Our analysis identified ransomware as the most common type of external attack carried against universities, and hacking for the purpose of personal gain as the most common form of internal attack. According to NCSC \cite{NCSC}, attackers regularly exploit weak passwords, lack of multi-factor authentication (MFA), and unpatched vulnerabilities in software; they also often have previous knowledge of user credentials, which further aids the attack \cite{NCSC}. Recently, attackers would have sought to engage in sabotaging backup or auditing devices, encrypting entire virtual servers, and using scripting environments (e.g., PowerShell) to easily deploy tooling or ransomware \cite{NCSC}. 

We now proceed to discuss the insider threat posed by students worldwide, including detailed recommendations from the findings in the present study. This is followed by a discussion of the overall impact and mitigation of other threats faced by universities, including ransomware, phishing and malware. Additional recommendations are also compiled and compared with the latest JISC report. We then conclude this section by discussing strengths and limitation of the work.

\subsection{Insider threat posed by students}
In the educational context, students are a particular case of insiders who have been given unique privileged access to systems. Because of this trust and access granted, insider threats may manifest as either malicious, complacent, or unintentional acts that negatively affect the organisation and its assets, including information technology (IT), facilities, networks, or data, and resulting in various types of harmful acts - e.g., fraud, theft, sabotage, or violence \cite{CISA}.

Our analysis depicted clear differences between school and university students in relation to motivation and types of attacks carried, with some similarities including cases of grade manipulation. Both categories of students may present a serious threat, requiring complex, proactive and prevention-focused mitigation efforts. Students should be guided to act in the interest and benefit of the organisation, following insider threat programs that 'deter, detect, and prevent people from wrongdoing' and when the harm has already been done, mitigate the impact(s) through appropriate management or enforcement actions' \cite{CISA}. Balancing focus, policy, processes, and messaging is important for universities to properly manage threats posed by students, which should be focused on defining who is posing the threat and in what capacity, including prioritising resources that need to be protected \cite{CISA}. This is in line with our findings showing school and university students may be motivated by different values, with the exception of grade manipulation, therefore calling for tailored responses to insider threat.

Moreover, a study of 462 university students found substantial variations across overall cyber security judgment levels, judgement levels on different aspects of cyber security, and rational judgement competency \cite{yan2018finding}. Namely, 104 (23\%) students showed the lowest correct judgements (the percentage correct below 50\%), constituting the most risky group and the true weakest links that deserve special attention, as opposed to the average group (73\%), showing the percentage correct between 56\% and 90\%, and the safe group (4\%) for which the percentage correct was above 90\% \cite{yan2018finding}. These three groups have been compared to students with disabilities, regular students, and gifted students in education \cite{yan2018finding}, however, further research is needed to understand how and why the risky group has the lowest judgment level. Moreover, the overall 65\% correct rate suggests students as a group have ``some cybersecurity judgment knowledge rather than knowing nothing about it or having solid knowledge'' \cite{yan2018finding}.

Similar to this individual diversity, the weakest links in judging various aspect of cyber security include a) failing to report a potential cyber security breach in the computer system (as illustrated by a scenario in which a manager chooses to reenter the missing confidential information instead), and b) underplaying security risks. In contrast, the least challenging scenario with the highest correct rate concerns dealing with popups, showing students may have experience in encountering popups daily and know the potential danger \cite{yan2018finding}. Finally, since students' rational judgment underperformed their intuitive judgment, the former which generally leads to better reasoning and decision making, this may indicate lack of strong scientific knowledge of cyber security stemming from inadequate education and training \cite{yan2018finding}. Universities should focus their resources on these three particular weakest links, providing special cyber security prevention and intervention programmes (i.e., formal or informal education and training to undergraduates) rather than lingering with the strongest links (i.e., dealing with popups).

Others found international students were significantly more susceptible to email scams than domestic students, as were first-year students compared to other years of study for reasons such as language barriers and differing experiences with cyber crime in the country of origin, respectively \citep{broadhurst2019phishing}. Greater training for international students and first-year students was proposed as a method to help mitigate cyber threats \citep{broadhurst2019phishing}.

Students were also unable to reliably detect spear-phishing when emails used a strategy of authority, as opposed to using social proof, even though ``the malicious link destinations were obviously unrelated to the content of the email'' \cite{butavicius2016breaching}. Moreover, those that could better weigh the consequences of immediate and future events were less likely to click on both phishing and spear-phishing links, meaning lower cognitive impulsivity may have overall protective effect against targeted, high effort attacks such as spear-phishing threats \cite{butavicius2016breaching}. This is linked to what is known as dual processing models of persuasion assuming analytical (`central') and heuristics (`peripheral') modes of information reasoning that are at odds \cite{xu}, and has direct training implications to defend against phishing attacks. Namely, since decision making styles can be modified \cite{pinillos2011}, training people to defend against phishing attacks could focus on activating `central’ mode processing when people are judging emails \cite{butavicius2016breaching}. However, more research is needed to confirm whether such intervention can improve phishing email discrimination \cite{butavicius2016breaching}.

Consequently, a one-sized fits all approach to training and education may not be an engaging way for students to increase their cyber security awareness and ultimately reduce the risk of human factors for universities. On the other hand, well-designed courses that follow internationally recognised frameworks could further improve the reception of cyber security awareness \cite{redman2020improving}.

\subsection{Impact and mitigation of other threats}

Ransomware attacks can have devastating operational, reputational, financial, and psychological effects, e.g., in terms of recovery time needed to reinstate critical services or becoming subject to intense media coverage and scrutiny. Reputational damage can be particularly problematic given over 50\% of university income in the sector is derived from tuition fees and 65\% of students would not apply to a university with a reputation for poor cyber security \cite{NCSC2}.

Moreover, once a cyber attack becomes publicly known, universities are often given high fines and further disciplinary actions to comply, as was in the case of a university which accrued a £120,000 ICO (Information Commissioner’s Office) fine \cite{ico2018greenwich}, and another which was given 144 ICO recommendations, including the recommendation to remove the chair of the university’s data protection privacy group \cite{ico2020warwick}.

There can be an impact on staff at targeted institutions which can include financial losses, loss of employment, emotional impacts, health problems, impact on family relationships, self blame and behavioural changes \cite{button2009}. Feelings of fear, anger, distrust and helplessness may be intertwined with guilt and shame felt by the victims, giving rise to significant short- and long-term effects of depression, panic attacks and post-traumatic stress disorder \cite{Bada}. For example, the cyber attack at the University of Greenwich and the University of Lancaster caused significant distress to staff and students alike, as 21,000 records were posted online \cite{ico2018greenwich} and fraudulent invoices were sent to undergraduate applicants \cite{lancaster2019cyber}.

To mitigate against ransomware, organisations should implement a ‘defence in depth’ strategy, including having effective vulnerability management and secure Remote Desktop Protocol (RDP) services in place, constraining scripting and macros, and exercising ransomware response regularly \cite{NCSC2}. Backups of important data are crucial for quick recovery in the event of disaster, and it is recommended that educational institutions should have at least three backup copies of important data, on at least two separate devices, of which at least one must be off-site \cite{DfE}.

In addition to detrimental effects of ransomware, other types of cyber crime impose their own set of challenges. For example, phishing attacks may result in ‘payments to a fraudulent recipient, the loss of a victim’s login credentials, or nefarious malware spreading throughout the university network’ \cite{NCSC2}, whereas malware ‘may be designed to enable the theft of information, provide the attacker with long-term access to a system, or render machines and data inaccessible, until a payment is made’ \cite{NCSC2}. Mitigation should include both good security awareness among staff and students, and strong access and authentication controls, e.g., via security-conscious policies \cite{NCSC2}. 

However, psychological traits and individual differences among computer system users can further explain vulnerabilities to cyber security attacks and crimes, as cognitive biases and impaired brain health can make individuals more susceptible to exploitation by cyber criminals \cite{norris2019}. For example, specific psychological vulnerabilities are being targeted in online communication to exploit human factors, e.g., by using time-limited responses to enact peripheral information processing and therefore limit the amount of deliberation \cite{button2009}. Risk taking and low self-control are also important personality factors predicting Internet fraud victimisation \cite{button2009}. It is therefore counterproductive to implement security countermeasures based on procedures and policies without considering human behaviours \cite{almansoori}.

\subsection{Additional recommendations}

The 2022 JISC posture survey suggested five equally important aspects to reduce and manage cyber security threats, including leadership, technical basics, awareness and training, certification, and cyber insurance \cite{JISC2022}. Our case studies reinforce these findings outlining that cyber security training should be made mandatory for both the staff and students, followed by regular refresher training. Moreover, communicating regularly with staff and students regarding cyber security helps to embed a cyber security culture. The majority of students appear not to be receiving cyber security training at universities. Such training is useful for society when students graduate and enter the working world, whilst compromise of student accounts can still lead to escalation by cyber threat actors. Sometimes postgraduates have access to some resources and databases that staff do, i.e. for teaching purposes. Therefore, postgraduates should be treated the same as staff members.

This obligatory training should be supplemented by a culture of directing people to appropriate resources instead of using links and attachments within emails. This is because credentials are usually stolen or malware is installed through clicking on such links or downloading attachments. If links and attachments were not included in emails, and it was normalised that university staff members, in particular IT and cyber security teams and those that hold senior positions, do not engage in such practice, this could potentially limit effectiveness of authority-type spear phishing. Cyber security policies already in place should be assessed using an international standard, such as the ISO 27000 series, to identify and fill any gaps, i.e. robust disaster recovery plans and data backup policies. An overarching information policy should bring together all the requirements, standards and best practices that apply to the handling of information to provide a robust governance framework for information management across the University. 

\subsection{Strengths and limitations}
Although a wealth of academic literature exists on the use and nature of cyber attacks, we believe this is the first review of cyber threats in the education sector, focusing on insider threat posed by students worldwide and offering advice and guidance as complied from identified sources in addition to our own recommendations. The novel timeline of key cyber attacks is provided with detailed temporal representation of the chain of key cyber attacks, aiding an understanding of how institutions were exploited and impacted by cyber crime, and supplemented with the list of attacks carried by students against their home universities, colleges, or schools to help frame the discussion of insider threat.

Despite these advantages, the work is not without its limitation. First, our sample size is small and based on published incidents. There are probably many more unpublished incidents and more data can help strengthen this research. Also, in some cases the data about specific characteristics of attack was not available to the authors. Lastly, the cyber attacks may have involved multiple attack elements, e.g. a ransomware attack may have involved phishing, and a DoS attack before the ransomware was activated, but our data does not gather this level of granular detail.

\section{Conclusion and future works}\label{Conclusions} The analysis presented in this paper is the first review of cyber threats in the education sector, providing a novel timeline of cyber attacks. Our analysis showed that many universities were hit by ransomware of which the leading causes were spam and phishing emails. Further, possible causes of internal attacks carried by students were also highlighted, the management of which should be focused on preventing hacking; however, additional tailored responses may be required as university, college, and school students appear to be motivated by different values, as well as the provision for awareness training two to three times per year and suitable technological protection measures to supplement.  Without this, criminals will continue to successfully target the human element within organisations, exploiting human psychology and error.

Future studies could build upon the timeline, including conducting surveys/interviews with institutions to find details of attacks they are willing to divulge that have not been covered in literature or news. We also plan to examine links between university security policies and their history of cyber crime attacks, as we expect that universities with weak policies in place may be more vulnerable to attacks. We will investigate whether this is indeed the case and whether a predictive model can be used to confirm such relationship. Finally, as different factors may be relevant to the reporting decision for different businesses, it would be interesting to further explore cyber crime reporting by universities, including better understanding of how financial and reputational risks may influence the decision to report cyber crime.

\newgeometry{left=1cm,right=1cm, bottom=3cm, top=2cm}

\begin{landscape}
	\scriptsize
	\newcounter{rownumbers}
	\newcommand\Rownumber{\stepcounter{rownumbers}\arabic{rownumbers}}
	\renewcommand*{\arraystretch}{1.3}

	\begin{longtable}{|p{0.4cm}|p{0.9cm}|p{1.6cm}|p{1.6cm}|p{11.0cm}|p{5.5cm}|p{0.6cm}|}
	\caption{A timeline of cyber attacks that have compromised UK universities.}
 \\
		\hline
		\textbf{ID} &\textbf{Date} & \textbf{Target} & \textbf{Perpetrator. Attack type. Impact. Motive} &  \textbf{Description} & \textbf{Impact} & \textbf{Ref} \\ 
		\endhead
		
		\endfoot
		\multicolumn{7}{p{21cm}}{Perpetrator key: U:University Student; H:Hacker; O:Other; S:High school student; $\Omega$:Unknown} \\
		\multicolumn{7}{p{21cm}}{Attack type key: D: Denial of service; E:Extortion; H:Hacking; I:SQL injection; K:Keylogger; M:Malware; P:Phishing (or smishing); R:Ransomware; $\Theta$:unknown} \\
		\multicolumn{7}{p{21cm}}{Impact key: C:Confidentiality; I:Integrity; A:Availability}\\
		\multicolumn{7}{p{21cm}}{Motive key: F:Financial; O:Other; P:Prank; G:Personal gain; $\Phi$:Unknown}\\
		\multicolumn{7}{r}{{\textit{Continued on next page}}} \\ 
		\endfoot

		\hline
\label{Table-CombinedCyberAttacks}
		\rownumber & 11 Feb 1988 &  Millions of internet users & S.D.A.P & Robert Tappan Morris, a graduate student from Cornell University, created a virus intended to be fun. The worm was launched from MIT. Morris was charged under the 1986 Computer Fraud and Abuse Act. in 2006, Morris became a tenured professor at MIT. & 6,000 of the then 60,000 computers connected to the internet were attacked, access made temporarily unavailable. &\cite{FBI2022}\\ \hline 

		\rownumber & 02 Feb, 2000 & eBay, Amazon, CNN, Dell, and Yahoo& S.D.A.P & A Canadian high school student referring to himself as MaffiaBoy launches a DDoS against eBay, Amazon, CNN, Dell, and Yahoo. Receives 8 months open custody.& \$1.2 Billion of damage& \cite{Blackhat2022}\\ \hline 
			
		\rownumber & 19 Jun 2008 & Tesoro High School, USA& S.H.I.G & 2 high school students hack school computer to change their own grades & charged with 69 felonies including 34 counts of altering their own grades and those of others & \cite{Vijayan2008}\\ \hline 
			

		\rownumber & 02 Oct 2012 & Multiple* & H.I.C.F & Team GhostShell -- a hacktivist group -- runs its’ Project Hellfire campaign. Uses SQL injection attacks and targets weak passwords to steal email addresses, passwords, IDs, and the names of staff and students from universities across the globe in its. & Data is made public in Project WestWind. Personal information and credentials of students and staff stolen and posted online. & \cite{info2020hacktivist, osborne2012ghostshell, info2020hacker} \\ 
		
		& & \multicolumn{4}{p{20cm}|}{* University of Cambridge; Imperial College London; University of Edinburgh; Manchester University; University of Nottingham; University of Sheffield.} & \\ \hline
		
		\rownumber & 14 Jun 2013 &  University of Purdue & S.H.I.G & Three students hacked into Purdue professors' university accounts and changed at least three dozen of their own grades & Personal grades altered & \cite{DailyMail2013}\\ \hline 
		
		\rownumber & 05 May 2014 & Miami-Dade County Public School & S.H.I.G & Student hacked into grade system and changed his own and 4 other student grades & Personal grades altered &\cite{Stosh2018}\\ \hline 
			
		\rownumber & 12 Jan 2015 & University of Queensland & U.H.I.G & Student hacked the University grade system and altered his grades & Charged with 14 offences which included forgery, impersonation, and using a restricted computer without consent & \cite{Wiggins2015}\\ \hline 
		
		\rownumber & 23 Apr 2015 & University of Birmingham & U.H.I.G &  Student used keylogging software to steal passwords, log onto examination systems, and up his grades, for example, one from 57 to 73\%. & Charged under the computer misuse act, jailed for four months & \cite{McCarthy2015}\\ \hline 
		
		\rownumber & 21 May 2015 & University of London & $\Omega$.D.A.$\Phi$ & DDoS attack launched just before students sat their finals. & Moodle inaccessible for over four hours. & \cite{leyden2015ddos, mendoza2015university, parr2015cyberattack} \\ \hline
		
		\rownumber & 07 Dec 2015 & Multiple universities (unspecified) & H.D.A.$\Phi$ & A DDoS attack was launched on the Janet network, which impacted the majority of UK universities. The cyber criminals adjusted their point of attack based on Jisc’s Twitter updates informing universities of their progress to stabilise the network. Attack lasted four days. & Eduroam Wi-Fi network disrupted (no Wi-Fi access); access to .ac.uk domains disrupted (websites down). & \cite{havergal2015cyber, kerr2015universities, williams2015universities} \\ \hline
		
		\rownumber & 15 Jan 2016 & University of Georgia & U.H.I.P & Student at Georgia tech hacks rival university website and alters content & Charged with computer trespass, asked to apologise and perform community service & \cite{Johnson2014}\\ \hline 
	
		\rownumber & 13 to 17 Jun 2016 & University of Greenwich & $\Omega$.H.C.$\Phi$ & Cyber criminal - claiming to be disgruntled former student - gain full control of the database structure and data.  & Data for 21,000 people were stolen and put on the dark web, including data on full names, contact details, and physical and mental health issues. University website defaced. Stolen files posted on dark web. Significant distress caused to students and staff members. ICO investigation reports serious breach and fines £120,000. & \cite{micklethwaite2016greenwich, oneill2021greenwich, curtis2015university} \\ \hline
		
		\rownumber & Dec 2016 & Multiple universities*. & H.I.C.F & Russian-speaking hacker Rasputin performs SQL injection attack on at least 60 global university and government organisation web sites. & Access credentials sold on Darkweb. Databases containing personal information compromised.  & \cite{cimpanu2017hacker, gundert2017russian, paganini2017russian} \\ 
		
		& & \multicolumn{4}{p{20cm}|}{* University of Cambridge; University of Oxford; University of Chester; University of Leeds; University of Glasgow; University of the Highlands and Islands; University of the West of England; University of Edinburgh.} & \\ \hline
		
		\rownumber & 11 Jan 2017 & University of Iowa & U.H.I.G & Student changed his grades more than 90 times in 21 months having gained access to the system through a keylogger which stole the user if and password of staff. & Charged with federal computer-hacking charges & \cite{Vaas2017}\\ \hline 

		\rownumber & 11 Aug 2017 & Singapore Management University & U.H.I.G & Student hacked into University computers, altered his final examination grade from D+ to B, final overall grade from B to A- and downgraded the grades of other students.&  Charged with the Computer Misuse and Cybersecurity Act and sentenced to 16 weeks jail &\cite{Elena2020}\\ \hline 
		
		\rownumber & Nov 2017 & University of South Wales & U.K.I.F & Student used USB devices to install key logging software on university computers, which he used to obtain login details of 35 university staff members over an 18 month period. Five systems were compromised, including the system used to store exam papers, coursework, and marking.  Stolen papers were sold to students.& Hayder receives a 20 month jail sentence & \cite{cluley2021foolish, dewey2021student, sabin2021former} \\ \hline
		
		\rownumber & 03 Dec 2017 & Tenafly High school (New Jersey) & S.H.I.G & High school student hacks into school student system, changes grades, and submits an Ivy league application & Personal grades altered & \cite{Cimpanu2017a}\\ \hline 
		
		\rownumber & 2017 & unnamed US university & $\Omega$.P.I.F & University paid \$1.9m into a cyber criminal's bank account in response to a fake invoice & Loss of revenue & \cite{JISC2020} \\ \hline
		
		\rownumber & 02 May 2018 & Various US organisations & U.H.C.$\Phi$ & Student faced extradition and charges under the US Computer Fraud and Abuse Act (CFAA) in the USA due to having hacked various organisations in the US including including the Federal Reserve, Nasa, the Department of Energy and the FBI. His extradition was blocked by UK courts. & Extradition requests filed but blocked by UK courts. Student, having previously studied at the University of Nottingham, then Glasgow, moved on to study at the University of Suffolk before becoming an advisor on computer security systems for Hacker House.&  \cite{Bowcott2018}\\ \hline 
		
		\rownumber & 14 May 2018& Ygnacio Valley High School in Concord& S.H.I.G & Student altered his own grades, and reducing the grades of others. & Charged with 14 felonies &\cite{Lapin2018}\\ \hline 
	
		\rownumber & May 2018 & Multiple universities (unspecified) & H.P,M.C.F & Spear-phishing emails sent as part of the STOLEN PENCIL campaign containing a link that would download a malicious Chrome extension, leading to credentials being stolen largely from biomedical engineers. & University login credentials compromised. IP exfiltrated. & \cite{lyngaas2018suspected, netscout2020stolen, uchill2018stolen} \\ \hline
		
		\rownumber & Aug 2018 & Multiple universities (unspecified) & H.P.C.F & Academics and students at multiple unspecified UK universities were targeted by the `Silent Librarian' campaign. Phishing emails, often spoofing a university’s library with stolen branding, would direct people to a login page on a false domain that harvested login credentials. After submitting credentials, victims are redirected to their actual university library page, often leaving no indication that information had just been phished. & University login credentials compromised; valuable research stolen. & \cite{cimpanu2018iranian, secureworks2018back, mitre2021silent} \\ \hline
		
		\rownumber & 11 Sep 2018 & University of Edinburgh & H.D.A.$\Phi$ & The university’s campus network was hit by a DDoS attack during its Freshers’ Week when hundreds of new undergraduates were starting their courses. & University’s website, wireless connections, and other student services down for over 24 hours. & \cite{driz2018most, muncaster2018edinburgh, uniedinburgh2018ddos} \\ \hline
		
		\rownumber & 2018 to 2019 & unnamed 144 US and 176 non US universities & H.P,M.C.F & Cybercrime group launches the \textit{silent librarian} attack & Theft of around \$3.4 billion worth of IP, 31.5 TB of data, and compromised 7,998 university accounts. & \cite{JISC2020} \\ \hline
		
		\rownumber & 03 Aug 2019 & Tufts University, USA & U.H.I.G & Student expelled on accusations of hacking the university grade system and altering her own grades. Notably, the charges were never proven or seemingly, properly investigated. &  Personal grades altered &\cite{Whittaker2019}\\ \hline 
		
		\rownumber & 05 Aug 2019 & Encore Junior and Senior High School for the Arts, USA& S.H.I.G & 15 year old student hacked into networks to alter his own grades. & Charged with computer intrusion. &\cite{Diaz2019}\\ \hline
		
		\rownumber & 06 Aug 2019 & Donald Trump& S.H.C.P & 2 Students hacked into Donald Trump’s tax records. & Charged under unauthorized computer access act & \cite{Stempel2019} \\ \hline 
		
		\rownumber & 09 Aug 2019 & Blackboard VLE & S.H.CI.$\Phi$ & US high school student, hacked Blackboard because `he was bored’. & Student discovers test grades, medical records, café balances, cryptographically hashed passwords, and photos. He reports the bugs to Blackboard who addressed the bugs imminently. Student was suspended for 2 days but no legal action was taken. &\cite{Greenberg2019}\\ \hline 
		
		\rownumber & 17 Dec 2019& High School in Maryland & S.H.C.$\Phi$ & A high school student hacks into school network. & Maintains access for over a year to a high school IP TV network and around 5,962 student records. Eventually the student disclosed himself to the school IT department. No sanctions were applied as the student disclosed himself.
		 & \cite{Wood2019, Samsel2019}\\ \hline 
		
		\rownumber & 2019 & The University of Warwick & H.H.CI.F & Cyber criminals exploited a remote-viewing software that had been installed by a staff member to gain access to the administrative network. An internal report was unable to identify what data had been stolen due to the state of cyber security protections that had been in place. & University suffers multiple data breaches; Sensitive information on students, staff, and volunteers for research studies stolen; ICO investigation that made 144 recommendations including 23 that were classed as urgent; ICO recommended the removal of the chair of the university’s data protection privacy group. & \cite{teiss2020warwick, martin2020warwick, muncaster2020cwarwick} \\ \hline
		
		\rownumber & 19 Jul 2019 & Lancaster University & H.P.C.F & The university was exploited by a phishing attack, although further details on the attack are unavailable. Names, addresses, phone numbers, and email addresses were stolen from undergraduate student applicant data records. Student records and ID documents were stolen from the student records system. & Fraudulent invoices demanding large amounts of money sent to some undergraduate applicants; distress to students affected. & \cite{ashford2019phishing, calderbank2019lancaster, lancaster2019cyber} \\ \hline
		
		\rownumber & 31 Jul 2019 & University of York & $\Omega$.$\Theta$.C.$\Phi$ & A “malicious data breach” was reported, although further details on the attack are unavailable. &  Perpetrator accesses the administrative records of 88 students and views private data. Further data on 4,440 students were downloaded. & \cite{bbc2019york, uniyork2019data, monrad2019universities} \\ \hline
		
		\rownumber & Aug 2019 & Multiple unnamed universities & H.PM.C.F & Second wave of the \textit{Silent Librarian} campaign targeted multiple unspecified UK universities once more, with minor variations to the social engineering techniques and infrastructure used in the previous year. & University login credentials compromised; valuable research stolen. & \cite{walsh2019phishy, proofpoint2019silent, odonnell2019silent} \\ \hline
		
		\rownumber & 2019 & Unnamed UK university & $\Omega$.$\Theta$.C.$\Phi$ & Cyber attack & Breach of around 4,100 student records. & \cite{JISC2020} \\ \hline
		
		\rownumber & 2019 & unnamed UK colleges and service providers & $\Omega$.$\Theta$.A.$\Phi$ & Cyber attack & Attack impairs system access for college and their staff. & \cite{JISC2020} \\ \hline

		\rownumber & 2019 & Maastricht University & H.R.A.F & Ransomware attack launched on University & University believed to have paid a ransom of around £230,000. Recovery effort involved multiple departments and teams. & \cite{JISC2020} \\ \hline

		\rownumber & May 2020 & Cloud software provider Blackbaud & H.R.A.F & Blackbaud provides services for a number of UK universities*, suffered a ransomware attack. They prevented clients being locked out of their own data and servers by paying the ransom. However, the cyber criminals extracted a copy of a subset of data. Some financial data and passwords were stored in cleartext. & Personal data of staff, students, alumni, and supporters stolen; bank account information compromised; user passwords compromised. & \cite{kelion2020blackbaud, hellard2020ablackbaud, blackbaud2020security} \\ 
		
		& & \multicolumn{4}{p{20cm}|}{* Impacted HEIs included: Birmingham, De Montfort, Strathclyde, Exeter, York; Oxford Brookes, Loughborough, Leeds, London, Reading, UCL, Oxford} & \\ \hline
		
		\rownumber & 11 May 2020 & The University of Edinburgh & H.H.A.F & The ARCHER supercomputer was infected with cryptocurrency mining malware. It was exploited through compromised SSH credentials. & Supercomputer used to mine cryptocurrency for criminals; supercomputer shut down for several days. & \cite{cimpanu2020supercomputers} \\ \hline
		
		\rownumber & 28 Aug 2020 & Northumbria University &  H.R.A.$\Phi$ & Cyber criminals infiltrate the university’s networks, steal data, and encrypt machines before demanding a ransom payment. & University's clearing service unable to take calls; exams postponed; campus closed for seven days; students unable to access their online student portal or other platforms. & \cite{cimpanu2020supercomputers, cyware2020cyberattackers, weston2020supercomputers} \\ \hline
		
		\rownumber & Aug 2020 & Unnamed UK college & $\Omega$.$\Theta$.A.$\Phi$ & Cyber attack launched on results day & Students cannot access results online, results must be emailed out & \cite{JISC2020} \\ \hline
		
		\rownumber & 03 Sep 2020& Florida school district & S.D.A.$\Phi$ & 16-Year-Old Arrested  after eight DDOS attacks on the School’s Online Learning Systems.& Charged with ``computer use in an attempt to defraud'' & \cite{PCMag2020}\\ \hline 
		
		\rownumber & 04 Sep 2020 & Newcastle University & H.R.A.F & DoppelPaymer steals university backup files, encrypts systems, and threatens to release the data unless ransom is paid. Data included a full list of students, their departments, courses, and student numbers, and home addresses, phone numbers, and personal email addresses for students. & Stolen files posted online; many university IT systems not operational for weeks, including relating to timetabling, building access, and smartcard issuing. & \cite{muncaster2020anewcastle, hellard2020chackers, merkel2020newcastle} \\ \hline
		
		\rownumber & Sep 2020 & Multiple universities* & H.PM.C.F & Third wave of ``Silent Librarian’’ campaign & University login credentials compromised; valuable research stolen. & \cite{malwarebytes2020silent, seals2020silent, arghire2020iran} \\
		
		& & \multicolumn{4}{p{20cm}|}{* Glasgow Caledonian University; University of Bristol; University of Cambridge; University of York; University of Kent; King’s College London; Queen Mary University of London; University of Lincoln} &\\ \hline
		
		\rownumber & 03 Oct 2020 & Nanyang Technical University & U.H.I.F & Student hacked into stored value card for hostel pre-paid SIMS, air conditioning, food, drinks and cigarettes. & Stole \$34,000 worth of items, and causing \$80,000 worth of damage. Tee was charged with the Computer Misuse and Cybersecurity Act and sentenced to 10 months jail &\cite{Tang2020}\\ \hline 
		
		\rownumber & 13 Dec 2020 & London Southbank University (LSBU) & $\Omega$.D.A.$\Phi$ & Cyber attack disables Moodle, student hub and other university systems. & Students unable to access Moodle, their student hub, recorded lectures, or academic resources for over a month; University-wide institutional login systems down. & \cite{shead2021southbank, lsbu2021cyber, buila2021secrecy} \\ \hline
		
		\rownumber & 13 Feb 2021 & The Chancellor, Masters and Scholars of the University of Oxford & H.$\Theta$.I.F & Cyber criminals exploit a poorly secured server in the Strubi labs to gain access to lab equipment, pumps and pressure tools. & Biochemical preparation machines compromised; Cyber criminals attempted to sell access to the lab’s systems. & \cite{brewster2021oxford, reuters2021oxford, osborne2021oxford} \\ \hline
		
		\rownumber & 24 Feb 2021 & Queen’s University Belfast (QUB) & $\Omega$.$\Theta$.A.$\Phi$ & Systems targeted in cyber attack, full extent does not appear to have been revealed, but institution disabled access for staff `as a precaution'. & Regular outages still occurred in some university systems three months after the incident, causing delays across the university; Health Trusts disconnected a number of services connected to QUB, and stopped accepting messages from QUB email addresses, disrupting communication between staff and students working on wards or other trust settings. & \cite{meredith2021queens, madden2021queens, doyle2021queens} \\ \hline
		
		\rownumber & 05 Mar 2021 & University of the Highlands and Islands (UHI) & H.R.A.F & Cyber criminals from eastern Europe or the Baltic region launched a ransomware attack using various malware components including Cobalt Strike and Ryuk. & Student and staff services disrupted, including the remote access platform to network drives, files and applications, and printing services; facilities closed for a day; systems such as administration and accounting still impacted several months on; system re-build required. & \cite{sullivan2021uhi, macleod2021uhi, corfield2021uhi} \\ \hline 
		
		\rownumber & 17 Mar 2021 & University of Northampton & H.R.A.F & A ransomware attack at the university caused serious server problems, with phone lines and IT systems taken down. & Six-day outage of IT, telephone, and server networks; students unable to access study materials for on-campus and online induction; access to online teaching disrupted; email systems unavailable. & \cite{marzouk2021northampton, jay2021northampton} \\ \hline
		
		\rownumber & 13 Apr 2021 & University of Portsmouth & H.R.A.F & Ransomware caused disruption to IT services, with technical disruption continuing over a couple of weeks. & University campus closed for over a week; start of new term delayed; key IT systems taken offline; staff and students unable to log onto network.  & \cite{fishwick2021portsmouth, muncaster2021portsmouth, bbc2021bportsmouth} \\ \hline
		
		\rownumber & 14 Apr 2021 & University of Hertfordshire & H.$\Theta$.A.$\Phi$ & The university experienced a cyber attack that impacts several systems including those in the cloud. & Large parts of the network are disconnected; access to cloud-based services, including Canvas, Office 365, MS Teams, and Zoom, blocked; all online classes cancelled; VPN, data storage, and email access all offline. & \cite{coker2020ransomware, speed2021hertford, scroxton2021hertfordshire} \\ \hline
		
		\rownumber & 14 May 2021 & Glasgow Caledonian University & H.$\Theta$.A.$\Phi$ & Cyber attack impacts university systems. & Still not clear what happened. However, the university suggested that systems were closed down as a precautionary measure. & \cite{harkins2021gcu} \\ \hline
		
		\rownumber & 09 Jun 2021 & Lay public & U.E.C.G & Teeside student, poses as Amazon tech support and cons an elderly woman out of £40,000. &  5 month jail sentence &\cite{ITV2021b}\\ \hline 
		
		\rownumber & 09 Oct 2021 & University of South Wales & U.H.I.F & Student hacks university system and makes £20k selling exam papers, housemate finds students willing to buy the exam papers. &  First student makes £20k in sales, jailed for 20 months, second student receives suspended 9 month sentence.
		 &\cite{ITV2021a}\\ \hline 
		
		\rownumber & 12 Oct 2021 & University of Sunderland & H.$\Theta$.A.$\Phi$ & A cyber attack affects university systems. & University website, IT systems -- including the university’s records systems -- and telephone lines disabled. Online classes cancelled,  processing of marks and awards delayed, staff members had difficulty accessing their emails. & \cite{knowles2021sunderland, tung2021sunderland, coker2021bsunderland} \\ \hline
		
		\rownumber & 15 Mar 2022 & Heriot-Watt University & H.M.A.$\Phi$ & The university was breached with a malware attack on its VPN, which led to on-premises infrastructure being taken down for over a week, including access to the VPN for remote login, Oracle R12 finance systems, staff shared areas, and some staff and student directories. & All users of systems had to be moved to alternative systems. Remote learning was hampered for over a week. & \cite{speed2022heriot, speirs2022heriot, pease2022heriot} \\ \hline
		
		\rownumber & 09 Sep 2022 &HS. Examinations department in Sri Lanka& S.H.I.G & An unnamed Information and Communication Technology student steals the results of 270,000 A-level students, and enables access to them through his own web portal. & Confidentiality breach &\cite{Weerasinghe2022}\\ \hline 
		
		\rownumber & 16 Sep 2022 & Escambia County School District, USA& O.$\Theta$.I.G & Daughter and mother - were accused of rigging the homecoming queen crown by using her mother's account to cast fake votes. The charges were challenged by daughter and later dropped. & Integrity breach &  \cite{Mann2022}\\ \hline 
	\end{longtable}

	\normalsize
	
\end{landscape}

\restoregeometry.

\bibliographystyle{elsarticle-num}

\bibliography{bibliography}

\begin{thebibliography}{100}
\expandafter\ifx\csname url\endcsname\relax
  \def\url#1{\texttt{#1}}\fi
\expandafter\ifx\csname urlprefix\endcsname\relax\def\urlprefix{URL }\fi
\expandafter\ifx\csname href\endcsname\relax
  \def\href#1#2{#2} \def\path#1{#1}\fi

\bibitem{hesa2021staff}
HESA,
  \href{https://www.hesa.ac.uk/news/19-01-2021/sb259-higher-education-staff-statistics}{Higher
  education staff statistics: Uk, 2019/20}, [Online]. [Accessed 22 August 2021]
  (2021).
\newline\urlprefix\url{https://www.hesa.ac.uk/news/19-01-2021/sb259-higher-education-staff-statistics}

\bibitem{hesa2021student}
HESA,
  \href{https://www.hesa.ac.uk/news/27-01-2021/sb258-higher-education-student-statistics/numbers}{Higher
  education student statistics: Uk, 2019/20 - student numbers and
  characteristics}, [Online]. [Accessed 22 August 2021] (2021d).
\newline\urlprefix\url{https://www.hesa.ac.uk/news/27-01-2021/sb258-higher-education-student-statistics/numbers}

\bibitem{lee2020security}
J.~Lee,
  \href{https://www.infosecurity-magazine.com/opinions/security-threats-universities/}{Security
  threats facing universities}, Infosecurity Magazine, [Online]. [Accessed 26
  January 2021] (2020).
\newline\urlprefix\url{https://www.infosecurity-magazine.com/opinions/security-threats-universities/}

\bibitem{bongiovanni2019least}
I.~Bongiovanni, The least secure places in the universe? a systematic
  literature review on information security management in higher education,
  Computers \& Security 86 (2019) 350--357.

\bibitem{Singar2020role}
A.~V. Singar, K.~B. Akhilesh, Role of cyber-security in higher education, Smart
  Technologies: Scope and Applications (2020) 249--264.

\bibitem{NCSC}
NCSC, The cyber threat to universities. assessing the cyber security threat to
  uk universities,
  https://www.ncsc.gov.uk/report/the-cyber-threat-to-universities, accessed:
  2022-12-28 (2019).

\bibitem{JISC2022}
JISC, Cyber impact 2022: The impact of cyber security incidents on the uk’s
  further and higher education and research sectors. observations, advice and
  questions to ask,
  https://repository.jisc.ac.uk/8732/1/cyber-impact-report-2022.pdf, accessed:
  2023-05-20 (2022).

\bibitem{csbs}
M.~.~S. Department~for Digital, Culture, Cyber security breaches survey 2022,
  https://www.gov.uk/government/statistics/cyber-security-breaches-survey-2022/cyber-security-breaches-survey-2022,
  accessed: 2023-01-16 (2022).

\bibitem{teiss2020warwick}
TEISS, \href{https://www.teiss.co.uk/warwick-university-data-breaches/}{Warwick
  university suffered multiple breaches due to poor security protocols},
  [Online]. [Accessed 09 December 2020] (2020).
\newline\urlprefix\url{https://www.teiss.co.uk/warwick-university-data-breaches/}

\bibitem{bbc2020blackbaud}
B.~News, \href{https://www.bbc.co.uk/news/technology-53516413}{Blackbaud hack:
  Universities lose data to ransomware attack}, [Online]. [Accessed 10 December
  2020] (2020).
\newline\urlprefix\url{https://www.bbc.co.uk/news/technology-53516413}

\bibitem{cost}
F.~N. Editor, Cyberattacks on the education industry cost on average over
  £620,000 per attack,
  https://www.fenews.co.uk/skills/wilful-ignorance-survey-reveals-despite-a-good-understanding-of-cyber-risk-office-workers-are-unwilling-to-adapt-their-behaviour/,
  accessed: 2023-01-16 (2021).

\bibitem{ncsc2020septalert}
NCSC,
  \href{https://www.ncsc.gov.uk/news/alert-targeted-ransomware-attacks-on-uk-education-sector}{Alert:
  Targeted ransomware attacks on the uk education sector by cyber criminals},
  [Online]. [Accessed 09 December 2020] (2020a).
\newline\urlprefix\url{https://www.ncsc.gov.uk/news/alert-targeted-ransomware-attacks-on-uk-education-sector}

\bibitem{JISC2020}
JISC, Cyber impact: The impact of cyber security incidents on the uk's further
  and higher education and research sectors,
  https://beta.jisc.ac.uk/reports/cyber-impact, accessed: 2022-12-28 (2020).

\bibitem{loeppky}
J.~Loeppky, Universities are fighting a cyber security war on multiple fronts,
  https://www.itpro.co.uk/security/cyber-security/368395/universities-are-fighting-a-cyber-security-war-on-multiple-fronts,
  accessed: 2023-01-16 (2022).

\bibitem{lallie2021cyber}
H.~S. Lallie, L.~A. Shepherd, J.~R. Nurse, A.~Erola, G.~Epiphaniou, C.~Maple,
  X.~Bellekens, Cyber security in the age of covid-19: A timeline and analysis
  of cyber-crime and cyber-attacks during the pandemic, Computers \& security
  105 (2021) 102248.

\bibitem{NCSC2021}
NCSC, Ncsc annual review 2021,
  https://www.ncsc.gov.uk/collection/ncsc-annual-review-2021, accessed:
  2023-05-19 (2021).

\bibitem{jones}
C.~Jones, Universities a 'huge target' for nation-state attackers, warns ncsc,
  \url{https://www.itpro.co.uk/cyber-warfare/34444/universities-a-huge-target-for-nation-state-attackers-warns-ncsc},
  accessed: 2023-01-16 (2019).

\bibitem{chapman2018severity}
J.~Chapman, A.~Chinnaswamy, A.~Garcia-Perez,
  \href{https://www.proquest.com/docview/2018924640/fulltextPDF/EE26A6CDE8174190PQ/1}{The
  severity of cyber attacks on education and research institutions: A function
  of their security posture}, in: 13th International Conference on Cyber
  Warfare and Security (ICCWS 2018), 08-09 March {2018}, Washington DC,
  Academic Conferences International, Reading, 2018, pp. 111--119, [Online].
  [Accessed 28 February 2021].
\newline\urlprefix\url{https://www.proquest.com/docview/2018924640/fulltextPDF/EE26A6CDE8174190PQ/1}

\bibitem{pavlova2020enhancing}
E.~Pavlova, Enhancing the organisational culture related to cyber security
  during the university digital transformation, Information \& Security 46~(3)
  (2020) 239--249.

\bibitem{Redscan}
Redscan, The state of cyber security across uk universities: An analysis of
  freedom of information requests, https://www.redscan.com/media/The-state-of-
  cyber-security-across-UK-universities-Redscan-report.pdf, accessed:
  2020-12-09 (2020).

\bibitem{yan2018finding}
Z.~Yan, T.~Robertson, R.~Yan, S.~Y. Park, S.~Bordoff, Q.~Chen, E.~Sprissler,
  Finding the weakest links in the weakest link: How well do undergraduate
  students make cybersecurity judgment?, Computers in Human Behavior 84 (2018)
  375--382.

\bibitem{butavicius2016breaching}
M.~Butavicius, K.~Parsons, M.~Pattinson, A.~McCormac, Breaching the human
  firewall: Social engineering in phishing and spear-phishing emails,
  ArXiv:1606.00887[Pre-Print] (2016).

\bibitem{zwilling2019deal}
M.~Zwilling, D.~Lesjak, S.~Natek, K.~Phusavat, P.~Anussornnitisarn, et~al.,
  \href{https://www.researchgate.net/publication/335137029_how_to_deal_with_the_awareness_of_cyber_hazards_and_security_in_higher_education}{How
  to deal with the awareness of cyber hazards and security in (higher)
  education}, in: Thriving on Future Education, Industry, Business and Society.
  Proceedings of the Makelearn and TIIM International Conference, 15-17 May
  {2019}, Piran, ToKnowPress, Lublin, 2019, pp. 433--439, [Online]. [Accessed
  22 December 2020].
\newline\urlprefix\url{https://www.researchgate.net/publication/335137029_how_to_deal_with_the_awareness_of_cyber_hazards_and_security_in_higher_education}

\bibitem{potgieter2019awareness}
P.~Potgieter, The awareness behaviour of students on cyber security awareness
  by using social media platforms: A case study at central university of
  technology, in: 4th International Conference on the Internet, Cyber Security
  and Information Systems, 31 October-01 November {2019}, Johannesburg, Kalpa
  Publications in Computing, Manchester, 2019, pp. 272--280.

\bibitem{zwilling2020cyber}
M.~Zwilling, G.~Klien, D.~Lesjak, {\L}.~Wiechetek, F.~Cetin, H.~N. Basim, Cyber
  security awareness, knowledge and behavior: A comparative study, Journal of
  Computer Information Systems (2020) 1--16.

\bibitem{lane2007information}
T.~Lane, Information security management in australian universities: An
  exploratory analysis, Ph.D. thesis, Queensland University of Technology,
  Queensland (Jan. 2007).

\bibitem{broadhurst2019phishing}
R.~Broadhurst, K.~Skinner, N.~Sifniotis, B.~Matamoros-Macias, Y.~Ipsen,
  Phishing and cybercrime risks in a university student community,
  International Journal of Cybersecurity Intelligence \& Cybercrime 2~(1)
  (2019) 4--23.

\bibitem{aljohani2020measuring}
W.~Aljohani, N.~Elfadil, Measuring cyber security awareness of students: A case
  study at fahad bin sultan university, International Journal of Computer
  Science and Mobile Computing 9~(6) (2020) 141--155.

\bibitem{redman2020improving}
S.~M. Redman, K.~J. Yaxley, K.~F. Joiner, Improving general undergraduate cyber
  security education: A responsibility for all universities?, Creative
  Education 11~(12) (2020) 2541--2558.

\bibitem{nfib2021cybercrime}
NFIB,
  \href{https://data.actionfraud.police.uk/cms/wp-content/uploads/2021/07/CYBER-Dashboard-Assessment-20-21.pdf}{Cyber
  crime trends}, [Online]. [Accessed 15 September 2021] (2021).
\newline\urlprefix\url{https://data.actionfraud.police.uk/cms/wp-content/uploads/2021/07/CYBER-Dashboard-Assessment-20-21.pdf}

\bibitem{Lagazio}
M.~Lagazio, N.~Sherif, M.~Cushman, A multi-level approach to understanding the
  impact of cyber crime on the financial sector, Computers and Security 45~(3)
  (2014) 58--74.

\bibitem{Kemp}
S.~Kemp, D.~Buil-Gil, F.~Miró-Llinares, N.~Lord, When do businesses report
  cybercrime? findings from a uk study, Criminology and Criminal Justice 0~(0)
  (2021) 17488958211062359.

\bibitem{redscan2020analysis}
Redscan,
  \href{https://www.redscan.com/media/The-state-of-cyber-security-across-UK-universities-Redscan-report.pdf}{The
  state of cyber security across UK universities: An analysis of Freedom of
  Information requests}, Redscan Cyber Security Limited, London, 2020,
  [Online]. [Accessed 09 December 2020].
\newline\urlprefix\url{https://www.redscan.com/media/The-state-of-cyber-security-across-UK-universities-Redscan-report.pdf}

\bibitem{chapman2019safe}
J.~Chapman, How safe is your data? cyber-security in higher education, HEPI
  Policy Note 12~(DOE-SLC-6903-1) (2019) 1--6.

\bibitem{torjesen2019universities}
I.~Torjesen, \href{https://www.bmj.com/content/365/bmj.l1633}{Universities'
  data systems can be hacked in under two hours, cybersecurity testing shows},
  BMJ (2019) article no: 1633 [no pagination][Online]. [Accessed 19 December
  2020].
\newline\urlprefix\url{https://www.bmj.com/content/365/bmj.l1633}

\bibitem{bagshaw2015unsw}
E.~Bagshaw,
  \href{https://www.smh.com.au/national/nsw/unsw-facebook-site-hacked-for-the-second-time-in-two-days-20150907-gjgjqq.html}{Unsw
  facebook site hacked for the second time in two days}, The Sydney Morning
  Herald, [Online]. [Accessed 15 December 2020] (2015).
\newline\urlprefix\url{https://www.smh.com.au/national/nsw/unsw-facebook-site-hacked-for-the-second-time-in-two-days-20150907-gjgjqq.html}

\bibitem{malwarebytes2020silent}
Malwarebytes,
  \href{https://blog.malwarebytes.com/malwarebytes-news/2020/10/silent-librarian-apt-phishing-attack/}{Silent
  librarian apt right on schedule for 20/21 academic year}, Malwarebytes Labs
  Blog, [Online]. [Accessed 12 August 2021] (Jul. 2021).
\newline\urlprefix\url{https://blog.malwarebytes.com/malwarebytes-news/2020/10/silent-librarian-apt-phishing-attack/}

\bibitem{tabansky2018iran}
L.~Tabansky, Iran's cybered warfare meets western cyber-insecurity, Iran's
  Cybered Warfare Meets Western Cyber-Insecurity (2018) 121--141.

\bibitem{seals2020silent}
T.~Seals,
  \href{https://threatpost.com/silent-librarian-school-research-stealing/160099/}{Silent
  librarian goes back to school with global research-stealing effort},
  Threatpost, [Online]. [Accessed 13 February 2021] (2020).
\newline\urlprefix\url{https://threatpost.com/silent-librarian-school-research-stealing/160099/}

\bibitem{zurkus2018stolen}
K.~Zurkus,
  \href{https://www.infosecurity-magazine.com/news/stolen-pencil-targets-academic/}{Stolen
  pencil targets academic institutions}, Infosecurity Magazine, [Online].
  [Accessed 10 December 2020] (2018).
\newline\urlprefix\url{https://www.infosecurity-magazine.com/news/stolen-pencil-targets-academic/}

\bibitem{mitre2020stolen}
T.~M. Corporation, \href{https://attack.mitre.org/groups/G0086/}{Stolen
  pencil}, [Online]. [Accessed 05 March 2021] (2020).
\newline\urlprefix\url{https://attack.mitre.org/groups/G0086/}

\bibitem{netscout2020stolen}
Netscout,
  \href{https://www.netscout.com/blog/asert/stolen-pencil-campaign-targets-academia}{Stolen
  pencil campaign targets academia}, Netscout ASERT Blog, [Online]. [Accessed
  05 March 2021] (Dec. 2018).
\newline\urlprefix\url{https://www.netscout.com/blog/asert/stolen-pencil-campaign-targets-academia}

\bibitem{netscout2022}
netscout, Ddos threat intelligence report: Unveiling the new threat landscape,
  https://www.netscout.com/threatreport/ddos-threat-intelligence-report/,
  accessed: 2023-05-19 (2022).

\bibitem{NCSCalert}
NCSC, Further ransomware attacks on the uk education sector by cyber criminals.
  alert, version 2.2,
  https://www.ncsc.gov.uk/files/NCSC-Alert-further-ransomware-attacks-UK-education-sector-20210604.pdf,
  accessed: 2023-05-19 (2021).

\bibitem{coker2020ransomware}
J.~Coker,
  \href{https://www.infosecurity-magazine.com/news/ulster-university-also-suffered/}{A
  third of uk unis hit by ransomware in last 10 years}, Infosecurity Magazine,
  [Online]. [Accessed 10 December 2020] (2020).
\newline\urlprefix\url{https://www.infosecurity-magazine.com/news/ulster-university-also-suffered/}

\bibitem{NCS2022}
H.~Government, National cyber strategy 2022. pioneering a cyber future with the
  whole of the uk,
  \url{https://assets.publishing.service.gov.uk/government/uploads/system/uploads/attachment_data/file/1053023/national-cyber-strategy-amend.pdf},
  accessed: 2023-05-19 (2022).

\bibitem{ibm2022}
{IBM Security}, \href{https://www.ibm.com/downloads/cas/3R8N1DZJ}{Cost of a
  data breach}, accessed: 2023-05-19 (2022).
\newline\urlprefix\url{https://www.ibm.com/downloads/cas/3R8N1DZJ}

\bibitem{loohuis2020maastricht}
K.~Loohuis,
  \href{https://www.computerweekly.com/news/252477997/Maastricht-University-pays-200000-to-Russian-hackers}{Maastricht
  university pays €200, 000 to russian hackers}, Computer Weekly, [Online].
  [Accessed 15 December 2020] (2020).
\newline\urlprefix\url{https://www.computerweekly.com/news/252477997/Maastricht-University-pays-200000-to-Russian-hackers}

\bibitem{gt}
D.~Wu, Ucsf pays hackers \$1.14m to recover encrypted data,
  https://www.govtech.com/security/ucsf-pays-hackers-1-14m-to-recover-encrypted-data.html,
  accessed: 2023-01-21 (2020).

\bibitem{hernandez2020regis}
E.~Hernandez,
  \href{https://www.denverpost.com/2020/01/28/regis-university-ransomware-cyberattack/}{Denver’s
  regis university paid ransom to “malicious actors” behind campus
  cyberattack}, The Denver Post, [Online]. [Accessed 09 February 2021] (2020).
\newline\urlprefix\url{https://www.denverpost.com/2020/01/28/regis-university-ransomware-cyberattack/}

\bibitem{NCSC2}
NCSC, Alert: Further ransomware attacks on the uk education sector by cyber
  criminals,
  https://www.ncsc.gov.uk/news/alert-targeted-ransomware-attacks-on-uk-education-sector,
  accessed: 2022-12-26 (June 2021).

\bibitem{chen2016islamic}
S.~Chen,
  \href{https://www.scmp.com/news/china/policies-politics/article/1902268/islamic-state-hackers-attack-top-tier-chinese}{`islamic
  state hackers' attack top tier chinese university’s website urging holy
  war}, South China Morning Post, [Online]. [Accessed 19 December 2020] (2016).
\newline\urlprefix\url{https://www.scmp.com/news/china/policies-politics/article/1902268/islamic-state-hackers-attack-top-tier-chinese}

\bibitem{sunday2017it}
S.~Standard,
  \href{https://www.sundaystandard.info/it-expert-reportedly-behind-hacking-of-ub-website/}{It
  expert reportedly behind hacking of ub website}, [Online]. [Accessed 19
  December 2020] (2017).
\newline\urlprefix\url{https://www.sundaystandard.info/it-expert-reportedly-behind-hacking-of-ub-website/}

\bibitem{qaa2021emerging}
QAA,
  \href{https://www.qaa.ac.uk/docs/qaa/guidance/emerging-cyber-security-threats.pdf}{Emerging
  cyber security threats to the integrity of uk teaching and learning},
  [Online]. [Accessed 06 August 2021] (2021).
\newline\urlprefix\url{https://www.qaa.ac.uk/docs/qaa/guidance/emerging-cyber-security-threats.pdf}

\bibitem{mckie2021australia}
A.~McKie,
  \href{https://www.timeshighereducation.com/news/australia-warns-uk-over-essay-mills-hacking-university-websites}{Australia
  warns uk over essay mills hacking university websites}, Times Higher
  Education, [Online]. [Accessed 20 April 2022] (2021).
\newline\urlprefix\url{https://www.timeshighereducation.com/news/australia-warns-uk-over-essay-mills-hacking-university-websites}

\bibitem{mcgowan2019china}
M.~McGowan,
  \href{https://www.theguardian.com/australia-news/2019/jun/06/china-behind-massive-australian-national-university-hack-intelligence-officials-say}{China
  behind massive australian national university hack, intelligence officials
  say}, The Guardian, [Online]. [Accessed 19 December 2020] (2019).
\newline\urlprefix\url{https://www.theguardian.com/australia-news/2019/jun/06/china-behind-massive-australian-national-university-hack-intelligence-officials-say}

\bibitem{korea2012nk}
T.~K. Times,
  \href{https://www.koreatimes.co.kr/www/nation/2021/08/113_103011.html}{Nk
  suspected of trying to hack into seoul university}, [Online]. [Accessed 19
  December 2020] (2012).
\newline\urlprefix\url{https://www.koreatimes.co.kr/www/nation/2021/08/113_103011.html}

\bibitem{ziv2021cyber}
A.~Ziv,
  \href{https://www.haaretz.com/israel-news/tech-news/.premium-cyberattack-on-israeli-university-data-being-erased-right-now-1.10119912}{Cyberattack
  hits israel’s bar ilan university: ‘data is being erased right now’},
  Haaretz, [Online]. [Accessed 23 August 2021] (2021).
\newline\urlprefix\url{https://www.haaretz.com/israel-news/tech-news/.premium-cyberattack-on-israeli-university-data-being-erased-right-now-1.10119912}

\bibitem{ehrlich2021new}
A.~B.~S. Ehrlich,
  \href{https://www.sentinelone.com/labs/new-version-of-apostle-ransomware-reemerges-in-targeted-attack-on-higher-education/}{New
  version of apostle ransomware reemerges in targeted attack on higher
  education}, Sentinel Labs Blog, [Online]. [Accessed 16 March 2022] (Sep.
  2021).
\newline\urlprefix\url{https://www.sentinelone.com/labs/new-version-of-apostle-ransomware-reemerges-in-targeted-attack-on-higher-education/}

\bibitem{microsoft}
Microsoft,
  \href{https://query.prod.cms.rt.microsoft.com/cms/api/am/binary/RE5bUvv?culture=en-us&country=us}{Microsoft
  digital defense report 2022. illuminating the threat landscape and empowering
  a digital defense}, accessed: 2023-05-19 (2022).
\newline\urlprefix\url{https://query.prod.cms.rt.microsoft.com/cms/api/am/binary/RE5bUvv?culture=en-us&country=us}

\bibitem{holmes2019giessen}
A.~Holmes,
  \href{https://www.businessinsider.com/university-giessen-hack-paper-passwords-germany-38000-students-2019-12?r=US&IR=T}{A
  university had to hand out paper passwords to 38, 000 students and staff
  after being hacked}, Business Insider, [Online]. [Accessed 09 February 2021]
  (2019).
\newline\urlprefix\url{https://www.businessinsider.com/university-giessen-hack-paper-passwords-germany-38000-students-2019-12?r=US&IR=T}

\bibitem{gilmore2016campus}
J.~Gilmore,
  \href{https://news.berkeley.edu/2016/02/26/campus-alerting-80000-individuals-to-cyberattack/}{Campus
  alerting 80, 000 individuals to cyberattack}, Berkeley News, [Online].
  [Accessed 09 February 2021] (2016).
\newline\urlprefix\url{https://news.berkeley.edu/2016/02/26/campus-alerting-80000-individuals-to-cyberattack/}

\bibitem{maranga2019emerging}
M.~J. Maranga, M.~Nelson, Emerging issues in cyber security for institutions of
  higher education, International Journal of Computer Science and Network 8~(4)
  (2019) 371--379.

\bibitem{fitzpatrick2017student}
E.~Fitzpatrick, N.~Riebe,
  \href{https://www.cbc.ca/news/canada/edmonton/student-charged-with-cyber-crimes-in-u-of-a-malware-breach-1.3922790}{Student
  charged with cyber crimes in u of a malware breach}, Canadian Broadcasting
  Corporation, [Online]. [Accessed 19 December 2020] (2017).
\newline\urlprefix\url{https://www.cbc.ca/news/canada/edmonton/student-charged-with-cyber-crimes-in-u-of-a-malware-breach-1.3922790}

\bibitem{bisson2018rutgers}
D.~Bisson,
  \href{https://www.tripwire.com/state-of-security/security-data-protection/man-ordered-to-pay-8-6-million-for-launching-ddos-attacks-against-rutgers-university/}{Man
  ordered to pay \$8.6 million for launching ddos attacks against rutgers
  university}, Tripwire, [Online]. [Accessed 09 February 2021] (2018).
\newline\urlprefix\url{https://www.tripwire.com/state-of-security/security-data-protection/man-ordered-to-pay-8-6-million-for-launching-ddos-attacks-against-rutgers-university/}

\bibitem{cheng2020individual}
C.~Cheng, L.~Chan, C.-l. Chau, Individual differences in susceptibility to
  cybercrime victimization and its psychological aftermath, Computers in Human
  Behavior 108 (2020).

\bibitem{Rogers}
M.~K. Rogers, A two-dimensional circumplex approach to the development of a
  hacker taxonomy, Digital investigation 3~(2) (2006) 97--102.

\bibitem{speed2021hertford}
R.~Speed,
  \href{https://www.theregister.com/2021/04/15/university_hertfordshire_cyber_attack/}{University
  of hertfordshire pulls the plug on, well, everything after cyber attack},
  [Online]. [Accessed 22 June 2021] (2021).
\newline\urlprefix\url{https://www.theregister.com/2021/04/15/university_hertfordshire_cyber_attack/}

\bibitem{scroxton2021hertfordshire}
A.~Scroxton,
  \href{https://www.computerweekly.com/news/252499376/University-of-Hertfordshire-is-latest-academic-cyber-attack-victim}{University
  of hertfordshire is latest academic cyber attack victim}, Computer Weekly,
  [Online]. [Accessed 27 March 2022] (2021).
\newline\urlprefix\url{https://www.computerweekly.com/news/252499376/University-of-Hertfordshire-is-latest-academic-cyber-attack-victim}

\bibitem{DoF2022}
D.~for Education, Leo graduate and postgraduate outcomes: 2019 to 2020,
  https://explore-education-statistics.service.gov.uk/find-statistics/leo-graduate-and-postgraduate-outcomes/2019-20,
  accessed: 2023-05-19 (2022).

\bibitem{CISA}
Cybersecurity, I.~S. Agency,
  \href{https://www.cisa.gov/sites/default/files/publications/Insider%20Threat%20Mitigation%20Guide_Final_508.pdf}{Insider
  threat mitigation guide}, Security Week, [Online]. [Accessed 28 December
  2022] (11 2020).
\newline\urlprefix\url{https://www.cisa.gov/sites/default/files/publications/Insider%20Threat%20Mitigation%20Guide_Final_508.pdf}

\bibitem{xu}
S.~K. Almansoori~A., Al-Emran~M., Dual process models of persuasion, The
  international encyclopedia of media effects 13~(9) (2017) 1--3.

\bibitem{pinillos2011}
N.~Pinillos, N.~Smith, G.~Nair, P.~Marchetto, , C.~Mun, Philosophy's new
  challenge: Experiments and intentional action, Mind and Language 26~(1)
  (2011) 115--139.

\bibitem{ico2018greenwich}
ICO,
  \href{https://ico.org.uk/about-the-ico/news-and-events/news-and-blogs/2018/05/the-university-of-greenwich-fined-120-000-by-information-commissioner-for-serious-security-breach/}{The
  university of greenwich fined £120, 000 by information commissioner for
  “serious” security breach}, [Online]. [Accessed 12 January 2021] (2018).
\newline\urlprefix\url{https://ico.org.uk/about-the-ico/news-and-events/news-and-blogs/2018/05/the-university-of-greenwich-fined-120-000-by-information-commissioner-for-serious-security-breach/}

\bibitem{ico2020warwick}
ICO,
  \href{https://ico.org.uk/media/action-weve-taken/audits-and-advisory-visits/2617436/university-of-warwick-audit-executive-summary-v1_2.pdf}{The
  university of warwick data protection audit report}, [Online]. [Accessed 12
  January 2021] (2020).
\newline\urlprefix\url{https://ico.org.uk/media/action-weve-taken/audits-and-advisory-visits/2617436/university-of-warwick-audit-executive-summary-v1_2.pdf}

\bibitem{button2009}
M.~Button, C.~Lewis, J.~Tapley, Fraud typologies and victims of fraud:
  literature review, accessed: 2023-05-19 (2009).

\bibitem{Bada}
M.~Bada, J.~R. Nurse, The social and psychological impact of cyberattacks, in:
  Emerging cyber threats and cognitive vulnerabilities, Elsevier, 2020, pp.
  73--92.

\bibitem{lancaster2019cyber}
L.~University, \href{https://www.lancaster.ac.uk/news/phishing-attack}{Cyber
  incident}, [Online]. [Accessed 27 March 2022] (2019).
\newline\urlprefix\url{https://www.lancaster.ac.uk/news/phishing-attack}

\bibitem{DfE}
D.~for Education, Meeting digital and technology standards in schools and
  colleges,
  https://www.gov.uk/guidance/meeting-digital-and-technology-standards-in-schools-and-colleges/cyber-security-standards-for-schools-and-colleges,
  accessed: 2022-12-28 (2022).

\bibitem{norris2019}
N.~G., B.~A, D.~D., The psychology of internet fraud victimisation: a
  systematic review, Journal of Police Criminal Psychology 34 (2019) 231--245.

\bibitem{almansoori}
S.~K. Almansoori~A., Al-Emran~M., Exploring the frontiers of cybersecurity
  behavior: A systematic review of studies and theories, Applied Sciences
  13~(9) (2023) 5700.

\bibitem{FBI2022}
FBI, \href{https://www.fbi.gov/history/famous-cases/morris-worm}{Morris worm}
  (2022).
\newline\urlprefix\url{https://www.fbi.gov/history/famous-cases/morris-worm}

\bibitem{Blackhat2022}
Blackhat,
  \href{https://www.blackhatethicalhacking.com/articles/hacking-stories/mafiaboy-the-hacker-who-took-down-the-internet/}{Mafiaboy,
  the hacker who took down the internet} (2022).
\newline\urlprefix\url{https://www.blackhatethicalhacking.com/articles/hacking-stories/mafiaboy-the-hacker-who-took-down-the-internet/}

\bibitem{Vijayan2008}
V.~Jaikumar,
  \href{https://www.computerworld.com/article/2534339/teen-faces-38-years-in-jail-for-grade-tampering-hack.html}{Teen
  faces 38 years in jail for grade-tampering hack} (2008).
\newline\urlprefix\url{https://www.computerworld.com/article/2534339/teen-faces-38-years-in-jail-for-grade-tampering-hack.html}

\bibitem{info2020hacktivist}
I.~Magazine,
  \href{https://www.infosecurity-magazine.com/news/hacktivist-campaign-targets-universities/}{Hacktivist
  campaign targets universities}, [Online]. [Accessed 15 March 2021] (2012a).
\newline\urlprefix\url{https://www.infosecurity-magazine.com/news/hacktivist-campaign-targets-universities/}

\bibitem{osborne2012ghostshell}
C.~Osborne,
  \href{https://www.zdnet.com/article/ghostshell-leaks-120000-records-from-top-100-universities/}{Ghostshell
  leaks 120, 000 records from top 100 universities}, ZDNet, [Online]. [Accessed
  27 March 2022] (2012).
\newline\urlprefix\url{https://www.zdnet.com/article/ghostshell-leaks-120000-records-from-top-100-universities/}

\bibitem{info2020hacker}
I.~Magazine,
  \href{https://www.infosecurity-magazine.com/news/hacker-collective-leaks-one-million-records-vows/}{Hacker
  collective leaks one million records vows 'hellfire'}, Infosecurity
  Magazine[Online]. [Accessed 15 March 2021] (2012).
\newline\urlprefix\url{https://www.infosecurity-magazine.com/news/hacker-collective-leaks-one-million-records-vows/}

\bibitem{DailyMail2013}
D.~Mail,
  \href{https://www.dailymail.co.uk/news/article-2341689/Purdue-University-engineering-students-hacked-professors-accounts-change-grades--boosting-A-A.html}{Purdue
  university engineering students 'hacked into professors' accounts to change
  their grades - even boosting an a to an a+'} (2013).
\newline\urlprefix\url{https://www.dailymail.co.uk/news/article-2341689/Purdue-University-engineering-students-hacked-professors-accounts-change-grades--boosting-A-A.html}

\bibitem{Stosh2018}
S.~Brandon,
  \href{https://freedomhacker.net/2014-04-high-school-student-arrested-for-hacking-report-card-system-and-changing-grades/}{High
  school student arrested for hacking report card system and changing grades}
  (2018).
\newline\urlprefix\url{https://freedomhacker.net/2014-04-high-school-student-arrested-for-hacking-report-card-system-and-changing-grades/}

\bibitem{Wiggins2015}
W.~Nick,
  \href{https://www.abc.net.au/news/2015-12-01/student-allegedly-caught-hacking-into-uq-computer/6990230}{Law
  student who allegedly hacked university of queensland to improve grades
  appears in court} (2015).
\newline\urlprefix\url{https://www.abc.net.au/news/2015-12-01/student-allegedly-caught-hacking-into-uq-computer/6990230}

\bibitem{McCarthy2015}
M.~Ross,
  \href{https://www.birminghammail.co.uk/news/midlands-news/university-birmingham-exam-cheat-jailed-9105937}{University
  of birmingham exam cheat jailed after hacking computers to boost marks}
  (2015).
\newline\urlprefix\url{https://www.birminghammail.co.uk/news/midlands-news/university-birmingham-exam-cheat-jailed-9105937}

\bibitem{leyden2015ddos}
J.~Leyden,
  \href{https://www.theregister.com/2015/05/22/university_of_london_ddos_attack/}{Ddos
  attack downs university of london learning platform}, The Register, [Online].
  [Accessed 10 December 2020] (2015).
\newline\urlprefix\url{https://www.theregister.com/2015/05/22/university_of_london_ddos_attack/}

\bibitem{mendoza2015university}
M.~Mendoza,
  \href{https://www.techtimes.com/articles/55166/20150525/university-of-london-learning-platform-suffers-from-ddos-attack.htm}{University
  of london learning platform suffers from ddos attack}, Tech Times, [Online].
  [Accessed 27 March 2022] (2015).
\newline\urlprefix\url{https://www.techtimes.com/articles/55166/20150525/university-of-london-learning-platform-suffers-from-ddos-attack.htm}

\bibitem{parr2015cyberattack}
C.~Parr,
  \href{https://www.timeshighereducation.com/news/cyberattack-shuts-down-virtual-learning-platform/2020378.article?page=0}{‘cyberattack’
  shuts down virtual learning platform}, Times Higher Education, [Online].
  [Accessed 27 March 2022] (2015).
\newline\urlprefix\url{https://www.timeshighereducation.com/news/cyberattack-shuts-down-virtual-learning-platform/2020378.article?page=0}

\bibitem{havergal2015cyber}
C.~Havergal,
  \href{https://www.timeshighereducation.com/news/cyber-attack-janet-network-disrupts-university-internet-connections}{Cyber
  attack on janet network disrupts university internet connections}, Times
  Higher Education, [Online]. [Accessed 10 December 2020] (2015).
\newline\urlprefix\url{https://www.timeshighereducation.com/news/cyber-attack-janet-network-disrupts-university-internet-connections}

\bibitem{kerr2015universities}
C.~Kerr,
  \href{https://www.huffingtonpost.co.uk/2015/12/08/universities-across-the-country-have-lost-the-internet-following-cyber-attack_n_8749308.html}{Universities
  across the country have lost internet following cyber attack}, The Huffington
  Post UK, [Online]. [Accessed 27 March 2022] (2015).
\newline\urlprefix\url{https://www.huffingtonpost.co.uk/2015/12/08/universities-across-the-country-have-lost-the-internet-following-cyber-attack_n_8749308.html}

\bibitem{williams2015universities}
R.~Williams,
  \href{https://www.telegraph.co.uk/technology/internet-security/12038962/Universities-across-the-country-lose-internet-connections-following-cyber-attack.html}{Universities
  across the country lose internet connections following cyber attack}, The
  Telegraph, [Online]. [Accessed 27 March 2022] (2015).
\newline\urlprefix\url{https://www.telegraph.co.uk/technology/internet-security/12038962/Universities-across-the-country-lose-internet-connections-following-cyber-attack.html}

\bibitem{Johnson2014}
J.~Joe,
  \href{https://eu.jacksonville.com/story/news/crime/2015/02/25/gerogia-tech-student-gets-pretrial-intervention-hack-uga-computers/15652332007/}{Gerogia
  tech student gets pretrial intervention in hack of uga computers} (2014).
\newline\urlprefix\url{https://eu.jacksonville.com/story/news/crime/2015/02/25/gerogia-tech-student-gets-pretrial-intervention-hack-uga-computers/15652332007/}

\bibitem{micklethwaite2016greenwich}
J.~Micklethwaite,
  \href{https://www.standard.co.uk/news/london/university-of-greenwich-website-hacked-by-disgruntled-former-student-a3274671.html}{University
  of greenwich website 'hacked by disgruntled former student'}, The Evening
  Standard, [Online]. [Accessed 10 December 2020] (2016).
\newline\urlprefix\url{https://www.standard.co.uk/news/london/university-of-greenwich-website-hacked-by-disgruntled-former-student-a3274671.html}

\bibitem{oneill2021greenwich}
P.~H. O'Neill,
  \href{https://www.dailydot.com/debug/university-greenwich-student-hack/}{Hacker
  breaches university of greenwich, exposes 21, 000 people’s data}, Daily
  Dot, [Online]. [Accessed 21 August 2021] (2021).
\newline\urlprefix\url{https://www.dailydot.com/debug/university-greenwich-student-hack/}

\bibitem{curtis2015university}
C.~Joe,
  \href{https://www.itpro.co.uk/data-breaches/31170/university-of-greenwich-fined-120000-for-breach-of-sensitive-data}{University
  of greenwich fined £120,000 for breach of sensitive data}, IT Pro, [Online].
  [Accessed 27 March 2022] (2018).
\newline\urlprefix\url{https://www.itpro.co.uk/data-breaches/31170/university-of-greenwich-fined-120000-for-breach-of-sensitive-data}

\bibitem{cimpanu2017hacker}
C.~Cimpanu,
  \href{https://www.bleepingcomputer.com/news/security/hacker-rasputin-breaches-over-60-universities-and-government-agencies/}{Hacker
  rasputin breaches over 60 universities and government agencies}, Bleeping
  Computer, [Online]. [Accessed 27 March 2022] (2017).
\newline\urlprefix\url{https://www.bleepingcomputer.com/news/security/hacker-rasputin-breaches-over-60-universities-and-government-agencies/}

\bibitem{gundert2017russian}
L.~Gundert,
  \href{https://www.recordedfuture.com/recent-rasputin-activity/}{Russian-speaking
  hacker sells sqli for unauthorized access to over 60 universities and
  government agencies}, [Online]. [Accessed 12 July 2021] (2017).
\newline\urlprefix\url{https://www.recordedfuture.com/recent-rasputin-activity/}

\bibitem{paganini2017russian}
P.~Paganini,
  \href{https://securityaffairs.co/wordpress/56312/hacking/russian-hacker-rasputin-attacks.html}{Russian
  hacker rasputin breaches over 60 universities and government agencies},
  Security Affairs, [Online]. [Accessed 27 March 2022] (2017).
\newline\urlprefix\url{https://securityaffairs.co/wordpress/56312/hacking/russian-hacker-rasputin-attacks.html}

\bibitem{Vaas2017}
V.~Lisa,
  \href{https://nakedsecurity.sophos.com/2017/11/01/student-charged-by-fbi-for-hacking-his-grades-more-than-90-times/}{Student
  charged by fbi for hacking his grades more than 90 times} (2017).
\newline\urlprefix\url{https://nakedsecurity.sophos.com/2017/11/01/student-charged-by-fbi-for-hacking-his-grades-more-than-90-times/}

\bibitem{Elena2020}
C.~Elena,
  \href{https://www.straitstimes.com/singapore/courts-crime/asean-scholar-at-smu-jailed-16-weeks-for-hacking-into-profs-computer-and}{Asean
  scholar at smu jailed 16 weeks for hacking into professor's computer and
  changing grades} (2020).
\newline\urlprefix\url{https://www.straitstimes.com/singapore/courts-crime/asean-scholar-at-smu-jailed-16-weeks-for-hacking-into-profs-computer-and}

\bibitem{cluley2021foolish}
G.~Cluley,
  \href{https://securityaffairs.co/wordpress/56312/hacking/russian-hacker-rasputin-attacks.html}{''foolish''
  university hacker jailed after selling exam papers to fellow students},
  Bitdefender, [Online]. [Accessed 27 March 2022] (2021).
\newline\urlprefix\url{https://securityaffairs.co/wordpress/56312/hacking/russian-hacker-rasputin-attacks.html}

\bibitem{dewey2021student}
P.~Dewey,
  \href{https://www.walesonline.co.uk/news/wales-news/student-hacked-university-system-sold-21528339}{Student
  hacked into university system and sold exam papers for £20, 000 profit},
  Wales Online, [Online]. [Accessed 16 March 2022] (2021).
\newline\urlprefix\url{https://www.walesonline.co.uk/news/wales-news/student-hacked-university-system-sold-21528339}

\bibitem{sabin2021former}
L.~Sabin,
  \href{https://www.independent.co.uk/news/uk/crime/hacker-prison-south-wales-university-exam-b1917451.html}{Former
  student hacker jailed for selling university exam answers}, Independent,
  [Online]. [Accessed 16 March 2022] (2021).
\newline\urlprefix\url{https://www.independent.co.uk/news/uk/crime/hacker-prison-south-wales-university-exam-b1917451.html}

\bibitem{Cimpanu2017a}
C.~Cimpanu,
  \href{https://www.bleepingcomputer.com/news/security/student-hacks-high-school-changes-grades-and-sends-college-applications/}{Student
  hacks high school, changes grades, and sends college applications} (Dec
  2017).
\newline\urlprefix\url{https://www.bleepingcomputer.com/news/security/student-hacks-high-school-changes-grades-and-sends-college-applications/}

\bibitem{Bowcott2018}
O.~Bowcott,
  \href{https://www.theguardian.com/law/2018/feb/05/hacking-suspect-lauri-love-wins-appeal-against-extradition-to-us}{Lauri
  love ruling 'sets precedent' for trying hacking suspects in uk} (2018).
\newline\urlprefix\url{https://www.theguardian.com/law/2018/feb/05/hacking-suspect-lauri-love-wins-appeal-against-extradition-to-us}

\bibitem{Lapin2018}
L.~Tamar,
  \href{https://nypost.com/2018/05/14/teens-hacks-school-to-change-grades-charged-with-14-felonies/}{Teen
  hacks school to change grades, charged with 14 felonies} (2018).
\newline\urlprefix\url{https://nypost.com/2018/05/14/teens-hacks-school-to-change-grades-charged-with-14-felonies/}

\bibitem{lyngaas2018suspected}
S.~Lyngaas,
  \href{https://www.cyberscoop.com/suspected-north-korean-hackers-target-universities-using-chrome-extension/}{Suspected
  north korean hackers target universities using chrome extension}, CyberScoop,
  [Online]. [Accessed 27 March 2022] (2018).
\newline\urlprefix\url{https://www.cyberscoop.com/suspected-north-korean-hackers-target-universities-using-chrome-extension/}

\bibitem{uchill2018stolen}
J.~Uchill,
  \href{https://www.axios.com/cybersecurity-academic-theft-stolen-pencil-north-korea-cdb0bbfe-2e6d-409d-9bb6-1358a002e4c0.html}{''stolen
  pencil'' data espionage campaign targets professors}, Axios, [Online].
  [Accessed 27 March 2022] (2018).
\newline\urlprefix\url{https://www.axios.com/cybersecurity-academic-theft-stolen-pencil-north-korea-cdb0bbfe-2e6d-409d-9bb6-1358a002e4c0.html}

\bibitem{cimpanu2018iranian}
C.~Cimpanu,
  \href{https://www.bleepingcomputer.com/news/security/iranian-hackers-charged-in-march-are-still-actively-phishing-universities/}{Iranian
  hackers charged in march are still actively phishing universities}, Bleeping
  Computer, [Online]. [Accessed 27 March 2022] (2018).
\newline\urlprefix\url{https://www.bleepingcomputer.com/news/security/iranian-hackers-charged-in-march-are-still-actively-phishing-universities/}

\bibitem{secureworks2018back}
Secureworks,
  \href{https://www.secureworks.com/blog/back-to-school-cobalt-dickens-targets-universities}{Back
  to school: Cobalt dickens targets universities}, [Online]. [Accessed 19
  December 2020] (2018).
\newline\urlprefix\url{https://www.secureworks.com/blog/back-to-school-cobalt-dickens-targets-universities}

\bibitem{mitre2021silent}
T.~M. Corporation, \href{https://attack.mitre.org/groups/G0122/}{Silent
  librarian}, [Online]. [Accessed 27 March 2022] (2021).
\newline\urlprefix\url{https://attack.mitre.org/groups/G0122/}

\bibitem{driz2018most}
S.~Driz,
  \href{https://www.drizgroup.com/driz_group_blog/most-universities-at-risk-of-ddos-attacks}{Most
  universities at risk of ddos attacks}, The Driz Group Cybersecurity Blog,
  [Online]. [Accessed 27 March 2022] (2018).
\newline\urlprefix\url{https://www.drizgroup.com/driz_group_blog/most-universities-at-risk-of-ddos-attacks}

\bibitem{muncaster2018edinburgh}
P.~Muncaster,
  \href{https://www.infosecurity-magazine.com/news/edinburgh-uni-hit-by-major-cyber/}{Edinburgh
  uni hit by major cyber-attack}, Infosecurity Magazine, [Online]. [Accessed 10
  December 2020] (2018).
\newline\urlprefix\url{https://www.infosecurity-magazine.com/news/edinburgh-uni-hit-by-major-cyber/}

\bibitem{uniedinburgh2018ddos}
T.~U. of~Edinburgh,
  \href{https://www.ed.ac.uk/infosec/information-security-updates/ddos-attack-10th-september-2018}{Ddos
  attack 10th september 2018}, [Online]. [Accessed 27 March 2022] (2018).
\newline\urlprefix\url{https://www.ed.ac.uk/infosec/information-security-updates/ddos-attack-10th-september-2018}

\bibitem{Whittaker2019}
Z.~Whittaker,
  \href{https://www.malwarebytes.com/blog/news/2022/09/when-students-hack-their-schools}{Tufts
  expelled a student for grade hacking. she claims innocence} (2019).
\newline\urlprefix\url{https://www.malwarebytes.com/blog/news/2022/09/when-students-hack-their-schools}

\bibitem{Diaz2019}
A.~Diaz,
  \href{https://www.latimes.com/california/story/2019-08-05/riverside-student-changing-grades-computer-hack}{Students
  are launching cyber-attacks against their schools} (2019).
\newline\urlprefix\url{https://www.latimes.com/california/story/2019-08-05/riverside-student-changing-grades-computer-hack}

\bibitem{Stempel2019}
S.~Jonathan,
  \href{https://www.reuters.com/article/instant-article/idINKCN1UW0KR}{College
  student who sought trump tax returns in cyber 'prank' to plead guilty}
  (2019).
\newline\urlprefix\url{https://www.reuters.com/article/instant-article/idINKCN1UW0KR}

\bibitem{Greenberg2019}
G.~Andy,
  \href{https://www.wired.com/story/teen-hacker-school-software-blackboard-follett/}{Student
  who hacked exams dept web arrested} (2019).
\newline\urlprefix\url{https://www.wired.com/story/teen-hacker-school-software-blackboard-follett/}

\bibitem{Wood2019}
C.~Wood,
  \href{https://edscoop.com/student-hacker-grabbed-personal-data-of-thousands-of-maryland-students/}{Student
  hacker grabbed personal data of thousands of maryland students} (Dec 2019).
\newline\urlprefix\url{https://edscoop.com/student-hacker-grabbed-personal-data-of-thousands-of-maryland-students/}

\bibitem{Samsel2019}
H.~Samsel,
  \href{https://campuslifesecurity.com/articles/2019/12/17/student-hacker-was-able-to-breach-thousands-of-student-accounts.aspx?admgarea=Topics&m=1}{Student
  hacker was able to breach thousands of student accounts in maryland district}
  (Dec 2019).
\newline\urlprefix\url{https://campuslifesecurity.com/articles/2019/12/17/student-hacker-was-able-to-breach-thousands-of-student-accounts.aspx?admgarea=Topics&m=1}

\bibitem{martin2020warwick}
A.~Martin,
  \href{https://news.sky.com/story/warwick-university-was-hacked-and-kept-breach-secret-from-students-and-staff-11978792}{Warwick
  university was hacked and kept breach secret from students and staff}, Sky
  News, [Online]. [Accessed 09 September 2020] (2020).
\newline\urlprefix\url{https://news.sky.com/story/warwick-university-was-hacked-and-kept-breach-secret-from-students-and-staff-11978792}

\bibitem{muncaster2020cwarwick}
P.~Muncaster,
  \href{https://www.infosecurity-magazine.com/news/warwick-uni-under-fire-reported/}{Warwick
  uni under fire after reported breach cover-up}, Infosecurity Magazine,
  [Online]. [Accessed 27 March 2022] (2020).
\newline\urlprefix\url{https://www.infosecurity-magazine.com/news/warwick-uni-under-fire-reported/}

\bibitem{ashford2019phishing}
W.~Ashford,
  \href{https://www.computerweekly.com/news/252467214/Phishing-attack-highlights-cyber-security-need-at-universities}{Phishing
  attack highlights cyber security need at universities}, Computer Weekly,
  [Online]. [Accessed 27 March 2022] (2019).
\newline\urlprefix\url{https://www.computerweekly.com/news/252467214/Phishing-attack-highlights-cyber-security-need-at-universities}

\bibitem{calderbank2019lancaster}
M.~Calderbank,
  \href{https://www.lep.co.uk/education/lancaster-university-hit-cyber-attack-hundreds-students-personal-data-stolen-650016}{Lancaster
  university hit by cyber attack, hundreds of students' personal data stolen},
  Lancashire Evening Post, [Online]. [Accessed 25 January 2021] (2019).
\newline\urlprefix\url{https://www.lep.co.uk/education/lancaster-university-hit-cyber-attack-hundreds-students-personal-data-stolen-650016}

\bibitem{bbc2019york}
B.~News,
  \href{https://www.bbc.co.uk/news/uk-england-york-north-yorkshire-49182179}{University
  of york students targeted in data hack}, [Online]. [Accessed 05 August 2021]
  (2019).
\newline\urlprefix\url{https://www.bbc.co.uk/news/uk-england-york-north-yorkshire-49182179}

\bibitem{uniyork2019data}
U.~of~York, \href{https://www.york.ac.uk/students/news/2019/data-breach/}{Data
  breach affecting a university system}, [Online]. [Accessed 27 March 2022]
  (2019).
\newline\urlprefix\url{https://www.york.ac.uk/students/news/2019/data-breach/}

\bibitem{monrad2019universities}
J.~Monrad,
  \href{https://www.infosecurity-magazine.com/opinions/universities-attackers/}{Universities
  fall into the cross hairs of cyber attackers}, Infosecurity Magazine,
  [Online]. [Accessed 25 January 2021] (2019).
\newline\urlprefix\url{https://www.infosecurity-magazine.com/opinions/universities-attackers/}

\bibitem{walsh2019phishy}
M.~Walsh,
  \href{https://www.proofpoint.com/us/threat-insight/post/seems-phishy-back-school-lures-target-university-students-and-staff}{Seems
  phishy: Back to school lures target university students and staff},
  Secureworks, [Online]. [Accessed 19 December 2020] (2019).
\newline\urlprefix\url{https://www.proofpoint.com/us/threat-insight/post/seems-phishy-back-school-lures-target-university-students-and-staff}

\bibitem{proofpoint2019silent}
Proofpoint,
  \href{https://www.proofpoint.com/us/threat-insight/post/threat-actor-profile-ta407-silent-librarian}{Threat
  actor profile: Ta407, the silent librarian}, Proofpoint, [Online]. [Accessed
  27 March 2022] (2019).
\newline\urlprefix\url{https://www.proofpoint.com/us/threat-insight/post/threat-actor-profile-ta407-silent-librarian}

\bibitem{odonnell2019silent}
L.~O'Donnell,
  \href{https://threatpost.com/silent-librarian-phishing-student-credentials/149249/}{Silent
  librarian retools phishing emails to hook student credentials}, Threatpost,
  [Online]. [Accessed 27 March 2022] (2019).
\newline\urlprefix\url{https://threatpost.com/silent-librarian-phishing-student-credentials/149249/}

\bibitem{kelion2020blackbaud}
L.~Kelion, \href{https://www.bbc.co.uk/news/technology-54370568}{Blackbaud:
  Bank details and passwords at risk in giant charities hack}, BBC News,
  [Online]. [Accessed 10 December 2020] (2020).
\newline\urlprefix\url{https://www.bbc.co.uk/news/technology-54370568}

\bibitem{hellard2020ablackbaud}
B.~Hellard,
  \href{https://www.itpro.co.uk/security/data-breaches/357314/blackbaud-admits-bank-data-lost-during-may-breach}{Blackbaud
  admits bank account details were lost in may data breach}, IT Pro, [Online].
  [Accessed 10 December 2020] (2020).
\newline\urlprefix\url{https://www.itpro.co.uk/security/data-breaches/357314/blackbaud-admits-bank-data-lost-during-may-breach}

\bibitem{blackbaud2020security}
Blackbaud, \href{https://www.blackbaud.com/securityincident}{Security
  incident}, [Online]. [Accessed 16 March 2022] (2020).
\newline\urlprefix\url{https://www.blackbaud.com/securityincident}

\bibitem{cimpanu2020supercomputers}
C.~Cimpanu,
  \href{https://www.zdnet.com/article/supercomputers-hacked-across-europe-to-mine-cryptocurrency/}{Supercomputers
  hacked across europe to mine cryptocurrency}, ZDNet, [Online]. [Accessed 21
  December 2020] (2020).
\newline\urlprefix\url{https://www.zdnet.com/article/supercomputers-hacked-across-europe-to-mine-cryptocurrency/}

\bibitem{cyware2020cyberattackers}
Cyware,
  \href{https://cyware.com/news/cyberattackers-targeting-supercomputers-british-supercomputer-archer-exploited-ddc26539}{Cyberattackers
  targeting supercomputers - british supercomputer archer exploited}, Cyware,
  [Online]. [Accessed 27 March 2022] (2020).
\newline\urlprefix\url{https://cyware.com/news/cyberattackers-targeting-supercomputers-british-supercomputer-archer-exploited-ddc26539}

\bibitem{weston2020supercomputers}
S.~Weston,
  \href{https://www.itpro.co.uk/security/malware/355677/uni-of-edinburgh-supercomputer-taken-down-by-cryptocurrency-mining-malware}{Supercomputers
  across europe taken down by crypto-mining malware}, IT Pro, [Online].
  [Accessed 27 March 2022] (2020).
\newline\urlprefix\url{https://www.itpro.co.uk/security/malware/355677/uni-of-edinburgh-supercomputer-taken-down-by-cryptocurrency-mining-malware}

\bibitem{PCMag2020}
P.~Mag,
  \href{https://uk.pcmag.com/education-reference/128478/16-year-old-arrested-for-cyberattacks-on-schools-online-learning-systems}{16-year-old
  arrested for cyberattacks on school's online learning systems} (2020).
\newline\urlprefix\url{https://uk.pcmag.com/education-reference/128478/16-year-old-arrested-for-cyberattacks-on-schools-online-learning-systems}

\bibitem{muncaster2020anewcastle}
P.~Muncaster,
  \href{https://www.infosecurity-magazine.com/news/newcastle-uni-ransomware-attack/}{Newcastle
  uni ransomware attack will “take weeks” to mitigate}, Infosecurity
  Magazine, [Online]. [Accessed 10 December 2020] (2020).
\newline\urlprefix\url{https://www.infosecurity-magazine.com/news/newcastle-uni-ransomware-attack/}

\bibitem{hellard2020chackers}
B.~Hellard,
  \href{https://www.itpro.co.uk/security/ransomware/357022/hackers-hold-newcastle-uni-student-data-to-ransom}{Hackers
  hold newcastle uni student data to ransom}, IT Pro, [Online]. [Accessed 27
  March 2022] (2020).
\newline\urlprefix\url{https://www.itpro.co.uk/security/ransomware/357022/hackers-hold-newcastle-uni-student-data-to-ransom}

\bibitem{merkel2020newcastle}
T.~Merkel,
  \href{https://thetab.com/uk/newcastle/2020/11/02/newcastle-students-data-including-home-addresses-leaked-on-dark-web-after-cyber-attack-52213}{Newcastle
  students’ data including home addresses leaked on dark web after cyber
  attack}, The Tab, [Online]. [Accessed 27 March 2022] (2020).
\newline\urlprefix\url{https://thetab.com/uk/newcastle/2020/11/02/newcastle-students-data-including-home-addresses-leaked-on-dark-web-after-cyber-attack-52213}

\bibitem{arghire2020iran}
I.~Arghire,
  \href{https://www.securityweek.com/iran-linked-silent-librarian-back-phishing-universities}{Iran-linked
  'silent librarian' back at phishing universities}, Security Week, [Online].
  [Accessed 27 March 2022] (2020).
\newline\urlprefix\url{https://www.securityweek.com/iran-linked-silent-librarian-back-phishing-universities}

\bibitem{Tang2020}
T.~Louisa,
  \href{https://www.todayonline.com/singapore/ntu-graduate-jailed-hacking-kopitiam-cards-free-items-causing-s80000-losses}{Ntu
  graduate jailed for hacking kopitiam cards for free items, causing \$80, 000
  in losses} (2020).
\newline\urlprefix\url{https://www.todayonline.com/singapore/ntu-graduate-jailed-hacking-kopitiam-cards-free-items-causing-s80000-losses}

\bibitem{shead2021southbank}
E.~Shead,
  \href{https://thetab.com/uk/london/2021/01/08/london-southbank-students-unable-to-access-online-learning-for-a-full-month-38309}{London
  southbank students unable to access online learning for a full month}, The
  Tab, [Online]. [Accessed 17 May 2021] (2021).
\newline\urlprefix\url{https://thetab.com/uk/london/2021/01/08/london-southbank-students-unable-to-access-online-learning-for-a-full-month-38309}

\bibitem{lsbu2021cyber}
L.~S.~B. University,
  \href{https://www.facebook.com/londonsouthbankuni/posts/10159051738682300}{Due
  to the recent cyber attack, the process for new students setting up passwords
  and accessing timetables looks a little different [...].}, Facebook,
  [Online]. [Accessed 27 March 2022] (2021).
\newline\urlprefix\url{https://www.facebook.com/londonsouthbankuni/posts/10159051738682300}

\bibitem{buila2021secrecy}
BUILA,
  \href{https://www.buila.ac.uk/news/2021/secrecy-and-cyberattacks-how-much-of-a-risk-is-information-sharing}{Secrecy
  and cyberattacks: how much of a risk is information sharing?}, BUILA,
  [Online]. [Accessed 27 March 2022] (2021).
\newline\urlprefix\url{https://www.buila.ac.uk/news/2021/secrecy-and-cyberattacks-how-much-of-a-risk-is-information-sharing}

\bibitem{brewster2021oxford}
T.~Brewster,
  \href{https://www.forbes.com/sites/thomasbrewster/2021/02/25/exclusive-hackers-break-into-biochemical-systems-at-oxford-uni-lab-studying-covid-19/}{Hackers
  break into ‘biochemical systems’ at oxford university lab studying
  covid-19}, Forbes, [Online]. [Accessed 17 May 2021] (2021).
\newline\urlprefix\url{https://www.forbes.com/sites/thomasbrewster/2021/02/25/exclusive-hackers-break-into-biochemical-systems-at-oxford-uni-lab-studying-covid-19/}

\bibitem{reuters2021oxford}
Reuters,
  \href{https://www.reuters.com/article/uk-health-coronavirus-britain-cyber-idUSKBN2AP2RP}{Oxford
  university says research not affected after expert flags covid lab hack},
  [Online]. [Accessed 17 May 2021] (2021).
\newline\urlprefix\url{https://www.reuters.com/article/uk-health-coronavirus-britain-cyber-idUSKBN2AP2RP}

\bibitem{osborne2021oxford}
C.~Osborne,
  \href{https://www.zdnet.com/article/oxford-university-biochemical-lab-involved-in-covid-19-research-targeted-by-hackers/}{Oxford
  university lab with covid-19 research links targeted by hackers}, ZDNet,
  [Online]. [Accessed 17 May 2021] (2021).
\newline\urlprefix\url{https://www.zdnet.com/article/oxford-university-biochemical-lab-involved-in-covid-19-research-targeted-by-hackers/}

\bibitem{meredith2021queens}
R.~Meredith, E.~Rosato,
  \href{https://www.bbc.co.uk/news/uk-northern-ireland-56287355}{Queen's
  university takes 'precautions' after cyber-attack attempt}, BBC News,
  [Online]. [Accessed 17 May 2021] (2021).
\newline\urlprefix\url{https://www.bbc.co.uk/news/uk-northern-ireland-56287355}

\bibitem{madden2021queens}
A.~Madden,
  \href{https://www.belfasttelegraph.co.uk/news/northern-ireland/belfast/health-service-disruption-fears-as-trusts-block-qub-emails-in-aftermath-of-attempted-cyber-attack-40502861.html}{Health
  service disruption fears as trusts block qub emails in aftermath of attempted
  cyber-attack}, Belfast Telegraph, [Online]. [Accessed 19 August 2021] (2021).
\newline\urlprefix\url{https://www.belfasttelegraph.co.uk/news/northern-ireland/belfast/health-service-disruption-fears-as-trusts-block-qub-emails-in-aftermath-of-attempted-cyber-attack-40502861.html}

\bibitem{doyle2021queens}
S.~Doyle,
  \href{https://www.irishnews.com/news/northernirelandnews/2021/03/06/news/queen-s-suspends-systems-after-attempted-cyber-attack-2245540/}{Queen's
  university suspends systems after attempted cyber attack}, The Irish News,
  [Online]. [Accessed 27 March 2022] (2021).
\newline\urlprefix\url{https://www.irishnews.com/news/northernirelandnews/2021/03/06/news/queen-s-suspends-systems-after-attempted-cyber-attack-2245540/}

\bibitem{sullivan2021uhi}
K.~Sullivan,
  \href{https://www.thetimes.co.uk/article/business-security-kit-used-for-cyberattack-on-highlands-university-250vfkl8k}{Business
  security kit used for cyberattack on highlands university}, The Times,
  [Online]. [Accessed 28 April 2021] (2021).
\newline\urlprefix\url{https://www.thetimes.co.uk/article/business-security-kit-used-for-cyberattack-on-highlands-university-250vfkl8k}

\bibitem{macleod2021uhi}
M.~Macleod,
  \href{https://www.stornowaygazette.co.uk/news/crime/college-cyber-attack-is-still-causing-problems-3222428}{College
  cyber-attack is still causing problems}, Stornoway Gazette, [Online].
  [Accessed 26 July 2021] (2021).
\newline\urlprefix\url{https://www.stornowaygazette.co.uk/news/crime/college-cyber-attack-is-still-causing-problems-3222428}

\bibitem{corfield2021uhi}
G.~Corfield,
  \href{https://www.theregister.com/2021/03/08/uni_highlands_islands_cyber_incident/}{University
  of the highlands and islands shuts down campuses as it deals with 'ongoing
  cyber incident'}, The Register, [Online]. [Accessed 27 March 2022] (2021).
\newline\urlprefix\url{https://www.theregister.com/2021/03/08/uni_highlands_islands_cyber_incident/}

\bibitem{marzouk2021northampton}
Z.~Marzouk,
  \href{https://www.itpro.co.uk/security/cyber-attacks/359003/university-of-northampton-hit-by-cyber-attack}{University
  of northampton hit by cyber attack}, IT Pro, [Online]. [Accessed 28 April
  2021] (2021).
\newline\urlprefix\url{https://www.itpro.co.uk/security/cyber-attacks/359003/university-of-northampton-hit-by-cyber-attack}

\bibitem{jay2021northampton}
J.~Jay,
  \href{https://www.teiss.co.uk/university-of-northampton-cyber-attack/}{Major
  cyber attack crippled northampton university’s it network for days}, TEISS,
  [Online]. [Accessed 28 April 2021] (2021).
\newline\urlprefix\url{https://www.teiss.co.uk/university-of-northampton-cyber-attack/}

\bibitem{fishwick2021portsmouth}
B.~Fishwick,
  \href{https://www.portsmouth.co.uk/education/university-of-portsmouth-closes-campus-due-to-ransomware-attack-on-it-services-causing-ongoing-disruption-3199053}{University
  of portsmouth closes campus due to 'ransomware attack' on it services causing
  'ongoing disruption'}, The News, [Online]. [Accessed 22 June 2021] (2021).
\newline\urlprefix\url{https://www.portsmouth.co.uk/education/university-of-portsmouth-closes-campus-due-to-ransomware-attack-on-it-services-causing-ongoing-disruption-3199053}

\bibitem{muncaster2021portsmouth}
P.~Muncaster,
  \href{https://www.infosecurity-magazine.com/news/campus-closed-portsmouth/}{Campus
  still closed as portsmouth university reels from suspected ransomware},
  Infosecurity Magazine, [Online]. [Accessed 22 June 2021] (2021).
\newline\urlprefix\url{https://www.infosecurity-magazine.com/news/campus-closed-portsmouth/}

\bibitem{bbc2021bportsmouth}
B.~News,
  \href{https://www.bbc.co.uk/news/uk-england-hampshire-56805248}{Portsmouth
  university 'cyber incident' reported to police}, [Online]. [Accessed 27 March
  2022] (2021).
\newline\urlprefix\url{https://www.bbc.co.uk/news/uk-england-hampshire-56805248}

\bibitem{harkins2021gcu}
D.~Harkins,
  \href{https://www.thetimes.co.uk/article/hackers-hit-it-systems-at-glasgow-caledonian-university-2h3lphgt0}{Hackers
  hit it systems at glasgow caledonian university}, The Times, [Online].
  [Accessed 26 July 2021] (2021).
\newline\urlprefix\url{https://www.thetimes.co.uk/article/hackers-hit-it-systems-at-glasgow-caledonian-university-2h3lphgt0}

\bibitem{ITV2021b}
ITV,
  \href{https://www.itv.com/news/tyne-tees/2021-09-06/teesside-university-cyber-security-student-posed-as-amazon-tech-support-in-scam}{Teesside
  university student studying cyber security posed as amazon tech support to
  con woman} (2021).
\newline\urlprefix\url{https://www.itv.com/news/tyne-tees/2021-09-06/teesside-university-cyber-security-student-posed-as-amazon-tech-support-in-scam}

\bibitem{ITV2021a}
ITV,
  \href{https://www.itv.com/news/wales/2021-09-10/student-hacker-who-made-20k-selling-exam-papers-jailed-for-20-months}{Student
  who hacked university system and made £20k selling exam papers jailed for 20
  months} (2021).
\newline\urlprefix\url{https://www.itv.com/news/wales/2021-09-10/student-hacker-who-made-20k-selling-exam-papers-jailed-for-20-months}

\bibitem{knowles2021sunderland}
T.~Knowles,
  \href{https://www.thetimes.co.uk/article/university-of-sunderland-it-system-still-disabled-cyberattack-83n5zwkz9}{University
  of sunderland’s it system still disabled a week after cyberattack},
  [Online]. [Accessed 16 March 2022] (2021).
\newline\urlprefix\url{https://www.thetimes.co.uk/article/university-of-sunderland-it-system-still-disabled-cyberattack-83n5zwkz9}

\bibitem{tung2021sunderland}
L.~Tung,
  \href{https://www.zdnet.com/article/university-still-recovering-from-major-cyberattack-that-disrupted-it-systems/}{University
  still recovering from major cyberattack that disrupted it systems}, ZDNet,
  [Online]. [Accessed 27 March 2022] (2021).
\newline\urlprefix\url{https://www.zdnet.com/article/university-still-recovering-from-major-cyberattack-that-disrupted-it-systems/}

\bibitem{coker2021bsunderland}
J.~Coker,
  \href{https://www.infosecurity-magazine.com/news/uni-sunderland-suspected-attack/}{University
  of sunderland hit by suspected cyber-attack}, Infosecurity Magazine,
  [Online]. [Accessed 27 March 2022] (2021).
\newline\urlprefix\url{https://www.infosecurity-magazine.com/news/uni-sunderland-suspected-attack/}

\bibitem{speed2022heriot}
R.~Speed, \href{https://www.theregister.com/2022/03/24/heriot_watt_outage/}{It
  outage at scotland's heriot-watt university enters second week}, The
  Register, [Online]. [Accessed 27 March 2022] (2022).
\newline\urlprefix\url{https://www.theregister.com/2022/03/24/heriot_watt_outage/}

\bibitem{speirs2022heriot}
K.~Speirs, J.~Delaney,
  \href{https://www.dailyrecord.co.uk/news/scottish-news/cyber-attack-scots-university-hampers-26549650}{Cyber
  attack at scots university hampers remote learning for over a week}, Daily
  Record, [Online]. [Accessed 27 March 2022] (2022).
\newline\urlprefix\url{https://www.dailyrecord.co.uk/news/scottish-news/cyber-attack-scots-university-hampers-26549650}

\bibitem{pease2022heriot}
V.~Pease,
  \href{https://news.stv.tv/east-central/police-investigate-cyber-attack-at-heriot-watt-university-in-edinburgh}{Police
  launch investigation after university hit by cyber attack}, STV, [Online].
  [Accessed 27 March 2022] (2022).
\newline\urlprefix\url{https://news.stv.tv/east-central/police-investigate-cyber-attack-at-heriot-watt-university-in-edinburgh}

\bibitem{Weerasinghe2022}
W.~G. Kumara,
  \href{https://www.dailynews.lk/2022/09/09/law-order/286885/student-who-hacked-exams-dept-web-arrested}{Student
  who hacked exams dept web arrested} (2022).
\newline\urlprefix\url{https://www.dailynews.lk/2022/09/09/law-order/286885/student-who-hacked-exams-dept-web-arrested}

\bibitem{Mann2022}
S.~Mann,
  \href{https://www.dailymail.co.uk/news/article-11219551/Florida-mean-girl-says-life-ruined-accused-rigging-homecoming-queen-election.html}{Florida
  ‘homecoming queen’ who says her life was ‘ruined’ after she and her
  vice-principal mom were accused of rigging election is launching a lawsuit
  against school and police officials claiming her civil rights were violated}
  (2022).
\newline\urlprefix\url{https://www.dailymail.co.uk/news/article-11219551/Florida-mean-girl-says-life-ruined-accused-rigging-homecoming-queen-election.html}

\end{thebibliography}

\end{document}